\documentclass[prd,preprint,preprintnumbers,nofootinbib,eqsecnum,superscriptaddress]{revtex4}

 \usepackage[dvips,final]{graphicx}
  \usepackage{amssymb}
   \usepackage{amsmath}
    \usepackage{amsfonts}
     \usepackage{epsfig}
      \usepackage{bm}

\usepackage{mathpazo}

\usepackage[section]{placeins}

\input epsf.tex
\def\desepsf(#1 width #2){\epsfxsize=#2 \epsfbox{#1}}

\usepackage[normalem]{ulem}

\usepackage{multirow}
\usepackage{ctable}
\usepackage{booktabs}
\usepackage{array}
\usepackage{tabularx}
\usepackage{xcolor}
\usepackage{pstricks}
\definecolor{fred}{rgb}{0.90053, 0.00369, 0.00159}  

\newcommand{\be}{\begin{eqnarray}}
\newcommand{\ee}{\end{eqnarray}}

\begin{document}

\title{The asymmetric intrinsic charm in the nucleon and its implications for the $\bm{D^{0}}$ production in the LHCb 
$\bm{p\!+\!\!^{20}\!N\!e}$ fixed-target experiment}

\author{Victor P.~Goncalves}
\email{barros@ufpel.edu.br}
\affiliation{Institute of Physics and Mathematics, Federal University of Pelotas, \\
  Postal Code 354,  96010-900, Pelotas, RS, Brazil}

\author{Rafa{\l} Maciu{\l}a}
\email{rafal.maciula@ifj.edu.pl}
\affiliation{Institute of Nuclear
Physics, Polish Academy of Sciences, ul. Radzikowskiego 152, PL-31-342 Krak{\'o}w, Poland}

\author{Antoni Szczurek\footnote{also at University of Rzesz\'ow, PL-35-959 Rzesz\'ow, Poland}}
\email{antoni.szczurek@ifj.edu.pl} 
\affiliation{Institute of Nuclear Physics, Polish Academy of Sciences, ul. Radzikowskiego 152, PL-31-342 Krak{\'o}w, Poland}


\begin{abstract}
Recent results indicate that charm quarks are intrinsic components of the proton wave function, and that the  charm and anticharm distributions for a given value of the Bjorken - $x$ variable can be different. In this paper, we will investigate the impact of this asymmetric intrinsic charm on the production of $D^0$ and ${\bar D}^0$ mesons for fixed target $p \!+ ^{20}\!\!Ne$ collisions at the LHCb. In our calculations, we include the contribution of the gluon-gluon fusion, gluon - charm and recombination processes and assume distinct models for the treatment of the intrinsic charm component. We demonstrate that the presence of an intrinsic charm improves the description of the current data for the  rapidity and transverse momentum distributions of $D$ mesons. However, such models are not able to describe  the LHCb data for the $D^0$-${\bar D}^0$ asymmetry at large transverse momentum, which point out that the description of the intrinsic charm needs to be improved and/or new effects should be taken into account in the production of heavy mesons at forward rapidities in fixed - target collisions.


\end{abstract} 

\maketitle

\section{Introduction}
The existence of an intrinsic charm (IC) component in the proton wave function has been proposed more than four decades ago and became a theme of intense theoretical and experimental investigation over tha last years (For a review see, e.g., Ref. \cite{Brodsky:2015fna}). In particular, the recent LHCb data for the $Z + c$ production \cite{LHCb:2021stx}, which was motivated by the propositions made in Refs. \cite{Bailas:2015jlc,Boettcher:2015sqn}, and the results obtained in Ref. \cite{Ball:2022qks}, where a global analysis of the parton distribution functions (PDFs) was performed, indicate that the presence of an intrinsic component improves the description of the current data. Such analyses have motivated more detailed studies focused on constraining the probability of finding an IC in the proton wave function ($P_{ic}$) and its phenomenological implications \cite{Carvalho:2017zge,Giannini:2018utr,Abdolmaleki:2019tbb,Maciula:2020dxv,Goncalves:2021yvw,Maciula:2022lzk}. In particular, in Ref.   \cite{Goncalves:2021yvw}, we have demonstrated that the measurements of high-energy atmospheric neutrinos by IceCube provide new limitations on $P_{ic}$.

\begin{figure}[t]
\begin{minipage}{0.32\textwidth}
  \centerline{\includegraphics[width=1.0\textwidth]{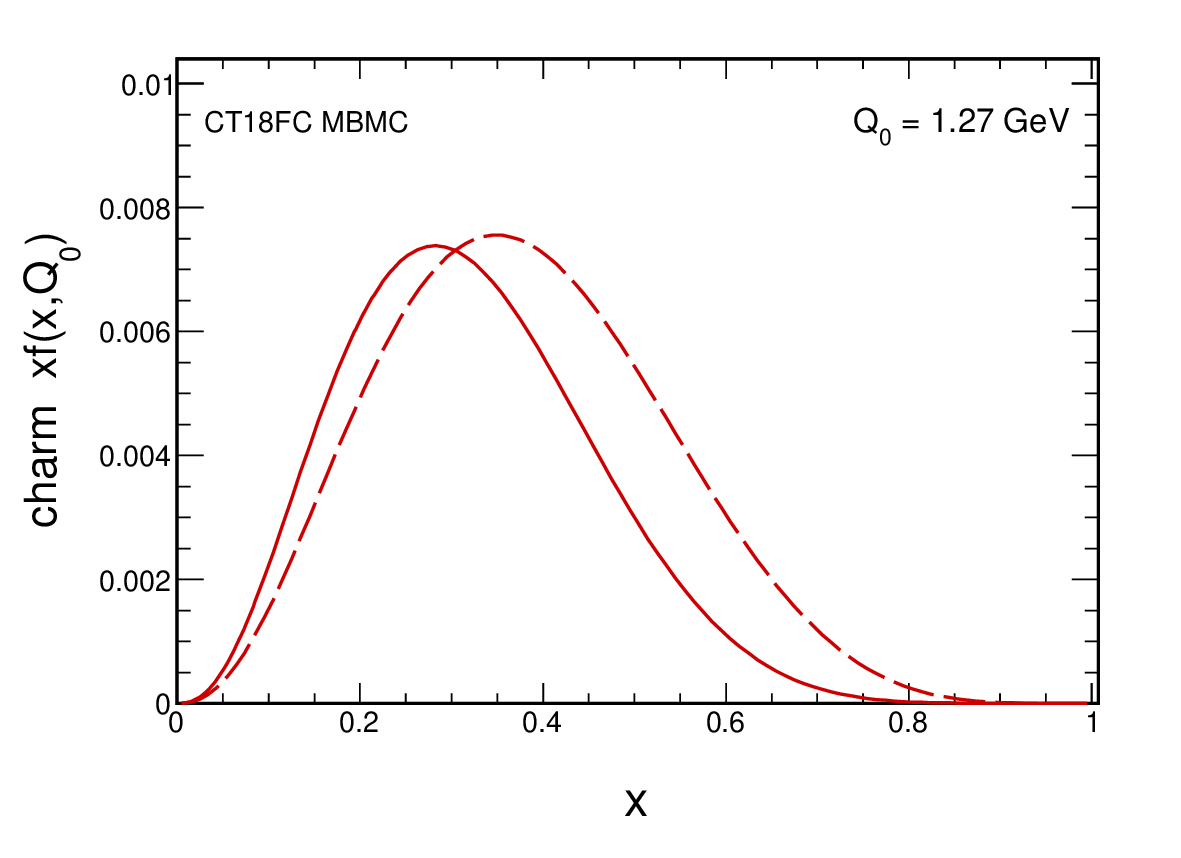}}
\end{minipage}
\begin{minipage}{0.32\textwidth}
  \centerline{\includegraphics[width=1.0\textwidth]{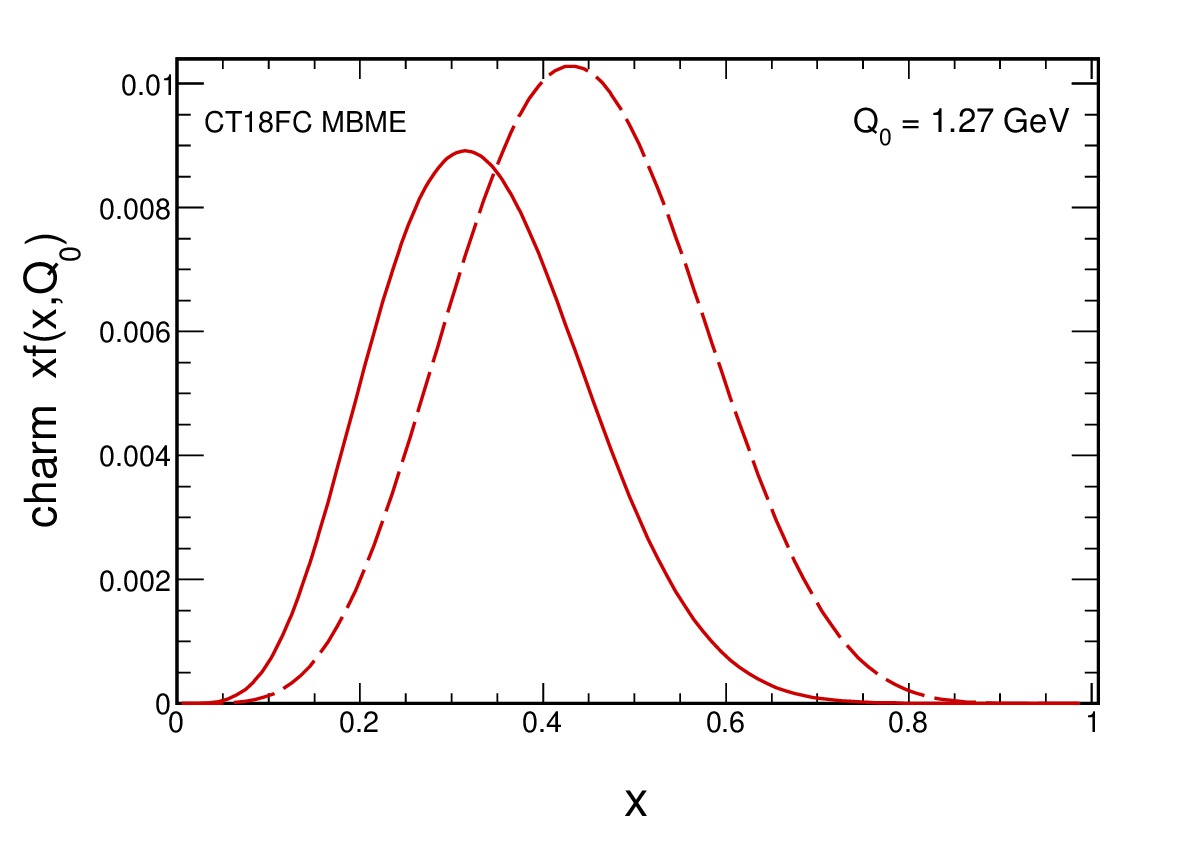}}
\end{minipage}
\begin{minipage}{0.32\textwidth}
  \centerline{\includegraphics[width=1.0\textwidth]{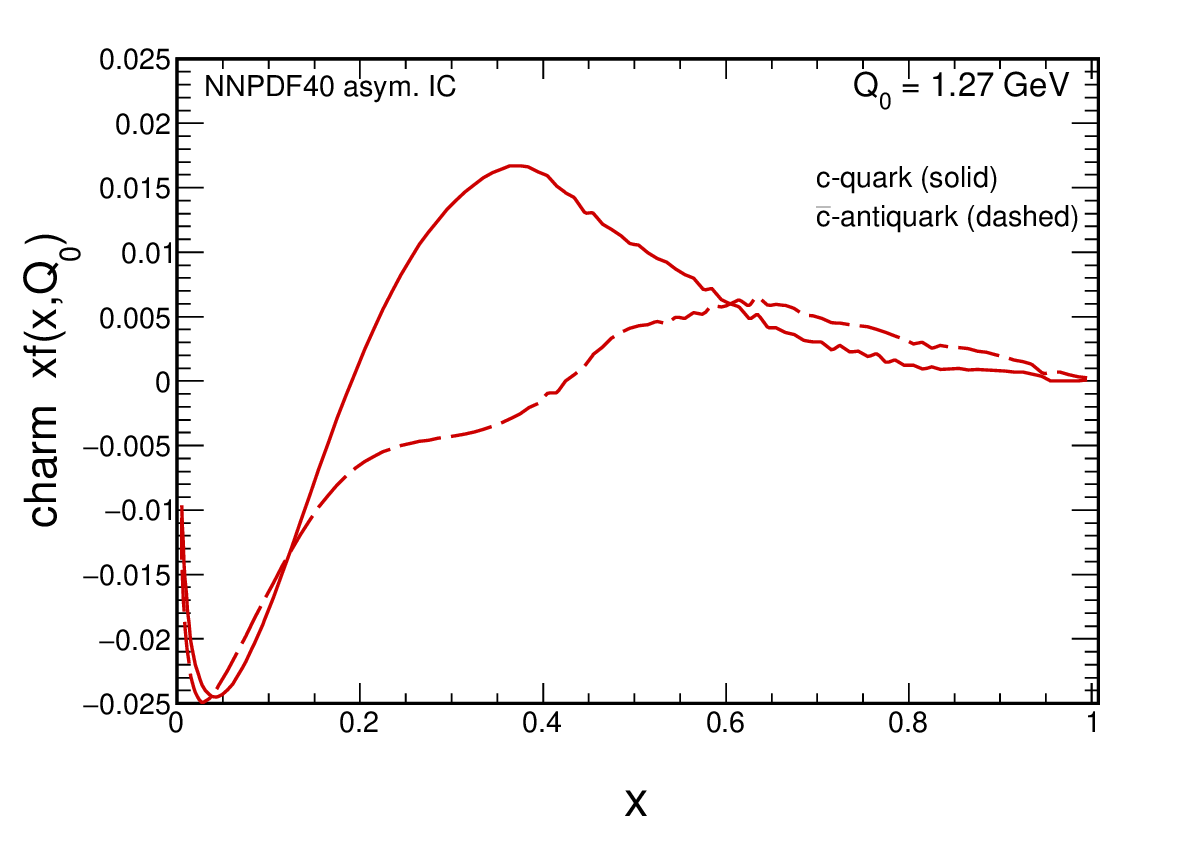}}
\end{minipage}
  \caption{
\small The distributions of charm quark (solid) and charm antiquark (dashed) as a function of $x$ at the initial scale $Q_{0}=1.27$ GeV for the CT18FC MBMC (left), CT18FC MBME (middle), and the NNPDF40 (right) PDFs. 
}
\label{fig:PDFs}
\end{figure}

A natural following question is, if a IC component is present, what is its dependence on the Bjorken - $x$ variable and if it is the same for charm and anticharm distributions.
The possibility of asymmetric intrinsic charm distributions,  $c(x) \neq \bar{c}(x)$, is predicted in the Meson Cloud models \cite{Navarra:1995rq,Paiva:1996dd,Hobbs:2013bia}, where the IC component comes from the proton fluctuation into an intermediate state composed by a charmed baryon plus a charmed meson. In contrast, in the model proposed by Brodsky,  Hoyer, Peterson and Sakai (BHPS) \cite{Brodsky:1980pb}, where this component is associated to higher Fock states, as e.g. the $|uudc\bar{c}\rangle$ state, one has $c(x) = \bar{c}(x)$. Such models have inspired the initial conditions assumed in the recent global analysis performed by the CTEQ - TEA   \cite{Guzzi:2022rca} and NNPDF  
\cite{NNPDF:2023tyk} groups. In particular, these groups provide parameterizations for an asymmetric intrinsic charm component that are consistent with the current experimental data, which are presented in Fig. \ref{fig:PDFs}, where we show the charm and anticharm distributions predicted by the NNPDF40 and CT18FC PDFs. For the CT18FC case, we present the results for two realizations  of the  meson-baryon model (MBM), based on confining (MBMC) and effective-mass (MBME) quark models, as discussed in detail in Ref.~\cite{Guzzi:2022rca}. 
The IC distributions in the NNPDF parametrization  are extracted without use of any model (e.g. BHPS and/or MBM), but instead by fitting charm to experimental data including both perturbative and non-perturbative components (the so-called fitted charm) and then removing the perturbatively generated part. 
One has that all these distinct sets predict a maximum for intermediate values of $x$, with the position and magnitude being different for charm and anticharm distributions. However, the results are strongly model dependent \footnote{It is important to emphasize that all these parametrizations satisfy the sum rule
\begin{eqnarray}
\int_0^1 dx [c(x,Q_0) - \bar{c}(x,Q_0)] = 0
\end{eqnarray}
at the initial scale $Q_0$.
}.

As the perturbative QCD evolution generates very small asymmetries in the charm and anticharm distributions, the measurement of a large asymmetry on the associated observables can be considered a signature of an intrinsic charm in the proton wave function. Considering that the IC contributes for large values of $x$, it is expected to modify the behavior of the observables at forward rapidities at the LHC.  Motivated by this expectation, in Ref. \cite{NNPDF:2023tyk} the authors proposed the study of charm asymmetries in the $Z+c$ production at forward rapidities in the LHC. In addition, they also have proposed the analysis of charm-tagged deep inelastic scattering at the future Electron - Ion Collider (EIC). In this paper, we will investigate  the impact of this asymmetric intrinsic charm on the production of $D^0$ and ${\bar D}^0$ mesons for fixed target $p \!+ ^{20}\!\!Ne$ collisions at the LHCb. In this process, characterized by smaller center - of - mass energies ($\sqrt{s_{NN}} = 68.5$ GeV), larger values of $x$ are accessed and the behavior of the differential distributions become sensitive to an IC charm component. Furthermore, the LHCb Collaboration has provided data for the $D^{0}$-${\bar D}^{0}$ 
asymmetry \cite{LHCb:2022cul}, which is the ideal observable to probe an asymmetric intrinsic charm. A shortcoming of this observable is that other effects, usually negligible at high energies, can also become relevant in the kinematical region probed in fixed - target collisions at the LHCb. For example, in the  study performed in Ref.~\cite{Maciula:2022otw}, two of us have demonstrated that the data for the differential distributions can only be described if the recombination process is included. However,  the LHCb asymmetry data is  only described for small meson transverse momenta. Here, we extend the previous study by analyzing whether the inclusion of an asymmetric intrinsic charm can improve the description of the current data.

This paper is organized as follows. In the next section, we will present a brief review of the theoretical framework used in our analysis. In particular, we will describe the formalism used to estimate the $D^0$ and  ${\bar D}^0$ production in fixed target $p\! + ^{20}\!\!Ne$ collisions at the LHCb. In section \ref{sec:results}, we present our results for the rapidity and transverse momentum differential distributions considering the contribution of the gluon - gluon, gluon - charm and recombination processes and distinct models for the asymmetric intrinsic charm. The predictions are compared with the current LHCb data. In addition, we present our results for the $D^{0}$-${\bar D}^{0}$ asymmetry. In section \ref{sec:conc} we summarize our results and main conclusions. Finally, in the Appendix, we discuss the improvement on the description of the fragmentation process assumed in our analysis and present, for completeness of our study, the predictions derived assuming a symmetric intrinsic charm.

\section{Theoretical framework}

Following our previous studies \cite{Maciula:2021orz,Maciula:2022otw}, here we also consider three different production mechanisms of charm, including:
a) the standard (and usually
considered as a leading) QCD mechanism of gluon-gluon fusion: $gg \to
c\bar c$; b) the gluon - charm channel, $gc \to gc$, which is enhanced by the presence of an intrinsic charm; and c) the recombination mechanism: $gq \to
Dc$. In what follows, we present a brief review of the main characteristics of these mechanisms and assumptions present in our calculations. 



\subsection{The standard QCD mechanism for charm production}


\begin{figure}[!h]
\centering
\begin{minipage}{0.3\textwidth}
  \centerline{\includegraphics[width=1.0\textwidth]{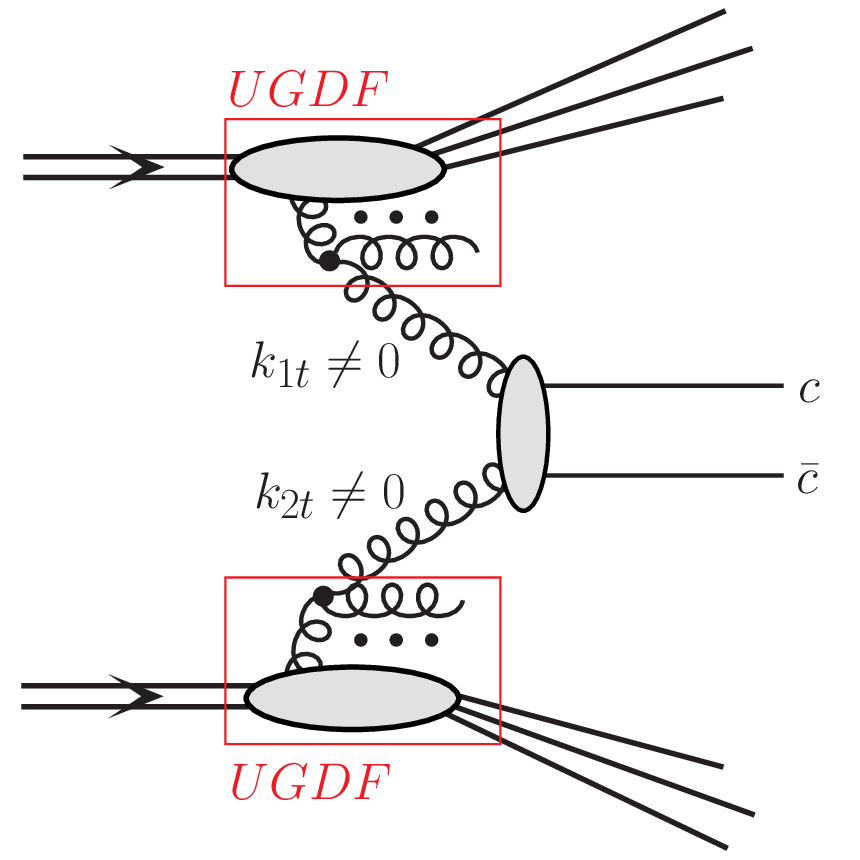}}
\end{minipage}
  \caption{
\small A diagram of the standard QCD mechanism of charm production in the $k_{T}$-factorization approach, driven by the fusion of two off-shell gluons.
}
\label{fig:diagramLO}
\end{figure}

The standard QCD mechanism for charm production is calculated within the $k_{T}$-factorization approach \cite{kTfactorization}, within the form
explored in the context of the LHCb fixed-target charm data in Ref.~\cite{Maciula:2020cfy}. In this framework, the transverse momenta
$k_{t}$'s (or virtualities) of both partons entering the hard process are taken into account, both in the matrix elements and in the parton distribution functions. Emission of the initial state partons is encoded in the transverse-momentum-dependent (unintegrated) PDFs (uPDFs).

In the case of charm flavor production within the so-called 0th flavor scheme, the parton-level cross-section is usually calculated via the $2\to 2$ mechanism of gluon-gluon fusion: $g^*g^* \to c\bar c$, with off-shell initial state gluons. Then the hadron-level differential cross-section for the $c \bar c$-pair production, reads:
\begin{eqnarray}\label{LO_kt-factorization} 
\frac{d \sigma(p p \to c \bar c \, X)}{d y_1 d y_2 d^2p_{1,t} d^2p_{2,t}} &=&
\int \frac{d^2 k_{1,t}}{\pi} \frac{d^2 k_{2,t}}{\pi}
\frac{1}{16 \pi^2 (x_1 x_2 s)^2} \; \overline{ | {\cal M}^{\mathrm{off-shell}}_{g^* g^* \to c \bar c} |^2}
 \\  
&& \times  \; \delta^{2} \left( \vec{k}_{1,t} + \vec{k}_{2,t} 
                 - \vec{p}_{1,t} - \vec{p}_{2,t} \right) \;
{\cal F}_g(x_1,k_{1,t}^2,\mu_{F}^2) \; {\cal F}_g(x_2,k_{2,t}^2,\mu_{F}^2) \; \nonumber ,   
\end{eqnarray}
where ${\cal F}_g(x_1,k_{1,t}^2,\mu_{F}^2)$ and ${\cal F}_g(x_2,k_{2,t}^2,\mu_{F}^2)$
are the gluon uPDFs for both colliding hadrons and ${\cal M}^{\mathrm{off-shell}}_{g^* g^* \to c \bar c}$ is the off-shell matrix element for the hard subprocess.
The gluon uPDF depends on gluon longitudinal momentum fraction $x$, transverse momentum
squared $k_t^2$ of the gluons entering the hard process, and in general also on a (factorization) scale of the hard process $\mu_{F}^2$. They must be evaluated at longitudinal momentum fractions 
$x_1 = \frac{m_{1,t}}{\sqrt{s}}\exp( y_1) + \frac{m_{2,t}}{\sqrt{s}}\exp( y_2)$, and $x_2 = \frac{m_{1,t}}{\sqrt{s}}\exp(-y_1) + \frac{m_{2,t}}{\sqrt{s}}\exp(-y_2)$, where $m_{i,t} = \sqrt{p_{i,t}^2 + m_c^2}$ is the quark/antiquark transverse mass. 

The kinematical regime probed in the fixed-target LHCb experiment corresponds to the region of longitudinal momentum fractions where, in principle, the Catani-Ciafaloni-Fiorani-Marchesini (CCFM) \cite{CCFM} evolution equation is legitimate for any pQCD theoretical calculations and could, in principle, be used to describe the dynamics of open charm meson production. In the numerical calculations below, we follow the conclusions from Refs.~\cite{Maciula:2020cfy,Maciula:2021orz} and apply the JH-2013-set2 gluon uPDFs \cite{Hautmann:2013tba} that are determined from high-precision DIS measurements. The CCFM gluon uPDFs were obtained within the 0th flavor scheme and are therefore compatible and can be safely applied in the model where the
$g^* g^* \to c \bar c$ mechanism is considered as leading. 

As a default set in the numerical calculations, we take the renormalization scale
$\mu^2 = \mu_{R}^{2} = \sum_{i=1}^{n} \frac{m^{2}_{it}}{n}$ (averaged
transverse mass of the given final state) and the charm quark mass
$m_{c}=1.5$ GeV. The strong-coupling constant $\alpha_{s}(\mu_{R}^{2})$
at next-to-next-to-leading-order is taken from the CT18FC PDFs
routines. The CCFM uPDFs here are taken at a rather untypical value of the factorization scale $\mu_{F}^2 = M_{c\bar c}^2 + P_{T}^{2}  $, where $M_{c\bar c}$ and $P_{T}$ are the $c\bar c$-invariant mass (or energy of the scattering subprocess) and the transverse momentum of $c\bar c$-pair (or the incoming off-shell gluon pair). This unusual definition has to be applied as a consequence of the CCFM evolution algorithm \cite{Hautmann:2013tba}.

\subsection{The gluon - charm mechanism}

\begin{figure}[t]
\centering
\begin{minipage}{0.4\textwidth}
  \centerline{\includegraphics[width=1.0\textwidth]{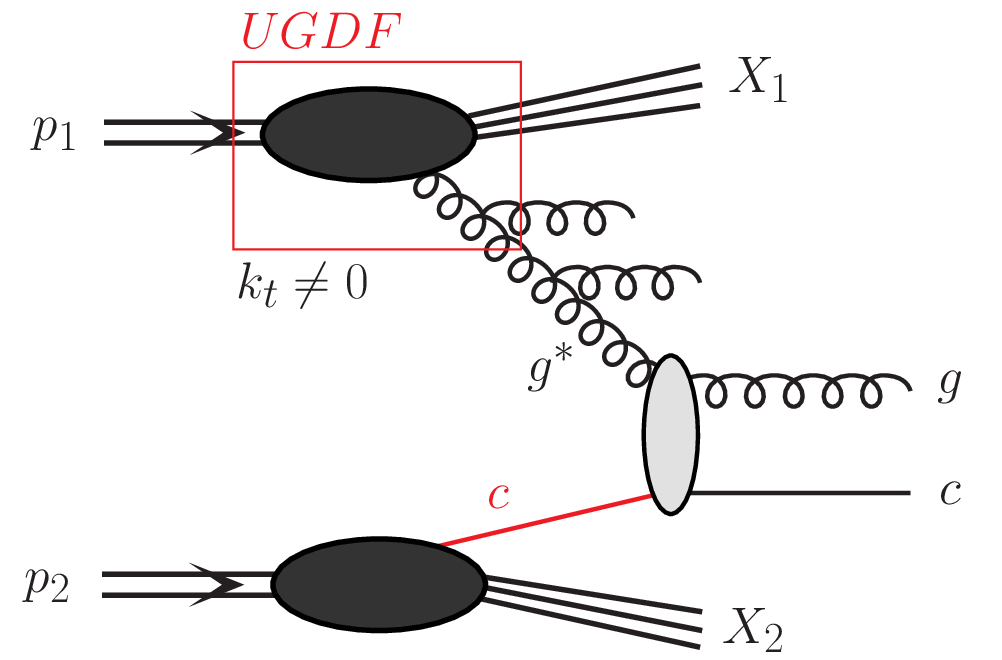}}
\end{minipage}
  \caption{
\small A diagrammatic representation of the gluon - charm mechanism for charm production within the hybrid model, with the off-shell gluon and the on-shell charm quark in the initial state.
}
\label{fig:diagramIC}
\end{figure}

The  contribution of the gluon - charm mechanism for the charm production cross-section is represented in 
Fig.~\ref{fig:diagramIC} and is obtained within the hybrid approach discussed in detail in Ref.~\cite{Maciula:2020dxv}. The LHCb
fixed-target configuration allows probing the charm cross-section in
the backward rapidity direction, where an asymmetric kinematical
configurations are preferred. Thus, in the basic $gc \to gc$ reaction, the gluon PDF and the  charm PDF are simultaneously examined at different longitudinal momentum fractions - rather intermediate for the gluon and large for the charm quark. The presence of an IC component implies an enhancement of this reaction, as demonstrated e.g. in Refs. \cite{Maciula:2020dxv,Goncalves:2021yvw}.

Within the asymmetric kinematic configuration, $x_1 \ll x_2$, the cross-section for the processes under consideration can be calculated in the so-called hybrid factorization model motivated by the work in Ref.~\cite{Deak:2009xt}. In this framework the small- or intermediate-$x$ gluon is taken to be off mass shell and 
the differential cross-section e.g. for $pp \to g c X$ via $g^* c \to g c$ mechanism reads:
\begin{eqnarray}
d \sigma_{pp \to gc X} = \int d^ 2 k_{t} \int \frac{dx_1}{x_1} \int dx_2 \;
{\cal F}_{g^{*}}(x_1, k_{t}^{2}, \mu^2) \; c(x_2, \mu^2) \; d\hat{\sigma}_{g^{*}c \to gc} \; ,
\end{eqnarray}
where ${\cal F}_{g^{*}}(x_1, k_{t}^{2}, \mu^2)$ is the unintegrated
gluon distribution in one proton and $c(x_2, \mu^2)$ a collinear PDF in
the second one. The hard partonic
cross-section, $d\hat{\sigma}_{g^{*}c \to gc}$, is obtained from a gauge invariant tree-level off-shell
amplitude. A derivation of the hybrid factorization from the dilute
limit of the Color Glass Condensate approach can be found e.g. in Ref.~\cite{Dumitru:2005gt} (see also Ref.~\cite{Kotko:2015ura}). The relevant cross-sections are calculated with the help of the \textsc{KaTie} Monte Carlo generator \cite{vanHameren:2016kkz}. There, the initial state quarks (including heavy quarks) can be treated as massless partons only.  

Working with minijets (jets with transverse momentum of the order of a few GeV) requires a phenomenologically motivated regularization of the cross-sections. Here we follow the minijet model \cite{Sjostrand:1987su} adopted e.g. in \textsc{Pythia} Monte Carlo generator, where a special suppression factor is introduced at the cross-section level \cite{Sjostrand:2014zea}:
\begin{equation}
F(p_t) = \frac{p_t^2}{ p_{T0}^2 + p_t^2 } \; 
\label{Phytia_formfactor}
\end{equation}
for each of the outgoing massless partons with transverse momentum $p_t$, where $p_{T0}$ is a free parameter of the form factor
that also enters as an argument of the strong coupling constant $\alpha_{S}(p_{T0}^2+\mu_{R}^{2})$. This suppression factor was originally proposed to remove the singularity present in the minijet cross-sections at leading-order  in the collinear approach. In the hybrid model (or in the $k_{T}$-factorization), the cross-sections are finite as long as $k_{T}> 0$, where $k_{T}$ is the transverse momentum of the incident off-shell parton. Within this approach, a treatment of the small-$k_{T}$ region in the construction of a given unintegrated parton density is crucial. Different models of uPDFs may lead to different behaviors of the cross-section at small minijet transverse momenta, but in any case the cross-sections should be finite. However, as it was shown in Ref.~\cite{Kotko:2016lej}, the internal $k_{T}$ cannot give a minijet suppression consistent with the minijet model and the related regularization seems to be necessary even in this framework.



\subsection{Recombination model and charm production}

\begin{figure}[t]
\centering
\includegraphics[width=5cm]{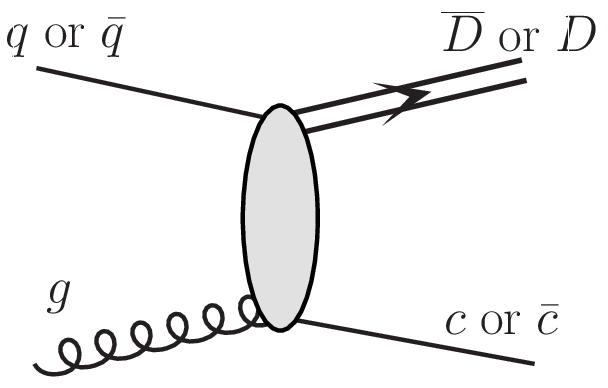}
\hskip+5mm
\includegraphics[width=5cm]{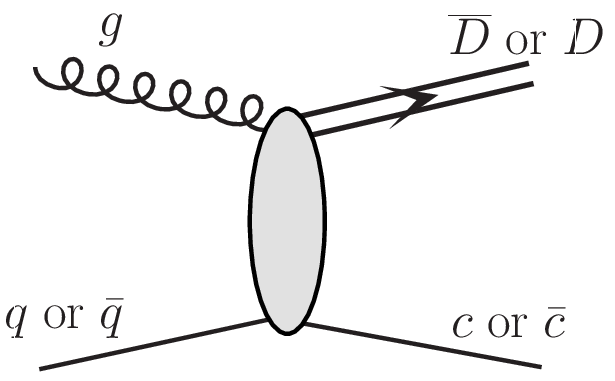}
\caption{Generic leading-order diagrams for $D$ meson production via 
the BJM recombination model.}
\label{fig:recombination_diagrams}
\end{figure}

The Braaten-Jia-Mechen (BJM) recombination mechanism \cite{BJM2002a,BJM2002b,BJM2002c} is illustrated 
in Fig.~\ref{fig:recombination_diagrams}. According to this approach, the differential cross-section for production of $D c$ final state reads:
\begin{eqnarray}
\frac{d\sigma}{d y_1 d y_2 d^2 p_{t}} = \frac{1}{16 \pi^2 {\hat s}^2}
&& [ x_1 q_1(x_1,\mu^2) \, x_2 g_2(x_2,\mu^2)
\overline{ | {\cal M}_{q g \to D c}(s,t,u)|^2} \nonumber \\
&+& x_1 g_1(x_1,\mu^2) \, x_2 q_2(x_2,\mu^2)
\overline{ | {\cal M}_{g q \to D c}(s,t,u)|^2} ]  \, .
\label{cross_section}
\end{eqnarray}
Above $y_1$ is rapidity of the $D$ meson and $y_2$ rapidity of 
the associated $c$ or $\bar c$. The matrix element squared in (\ref{cross_section}) can be written as
\begin{equation}
\overline{ | {\cal M}_{q g \to D c}(s,t,u)|^2} =
\overline{ | {\cal M}_{q g \to ({\bar c} q)^n c} |^2} \cdot \rho \; ,
\label{rho_definition}
\end{equation}
where $n$ enumerates quantum numbers of the ${\bar c} q$ system
$n \equiv ^{2J+1}L$ and $\rho$ can be interpreted as a probability to form the meson.
The parameter cannot be directly calculated and has to be extracted from the data. 
As we have shown in Ref.~\cite{Maciula:2022otw}, recent data on charm production asymmetry in fixed-target LHCb experiment
can be very useful in this context and help to adjust the precise number. 
For example, the asymmetries observed in photoproduction can be explained with $\rho$ = 0.15
\cite{BJM2002c}. Our rough estimation of the parameter from the charm data under consideration here will be also presented, when discussing numerical results. 

Within the recombination mechanism, we include two different ingredients: a) the direct recombination and the fragmentation of $c$-quarks or $\bar{c}$-antiquarks accompanying directly produced
$D$-mesons or $\overline{D}$-antimesons, e.g.:
\begin{eqnarray}
d \sigma [q g \to \overline{D}_{\mathrm{direct}} + D_{\mathrm{frag.}}] &=& d \sigma [q g \to \overline{D} + c] 
\otimes F^{\mathrm{frag.}}_{c \to D} \; ,
\label{fragmentation}
\end{eqnarray}
where $F^{\mathrm{frag.}}_{c \to D}$ is the relevant fragmentation function.
How the convolution $\otimes$ is understood
is explained in Ref.~\cite{Maciula:2019iak}.

\section{Results}
\label{sec:results}
In this section, we will present our predictions for the rapidity and transverse momentum distributions associated with the production of $D^0$ and ${\bar D}^0$ mesons in fixed target $p + ^{20}Ne$ collisions at the LHCb. The contribution of the three mechanisms discussed above will be estimated. 
In our  numerical calculations, the intrinsic charm PDFs will be taken at the initial scale $m_{c} = 1.3$ GeV, so the perturbative charm contribution is intentionally not taken into account. This makes the calculation compatible with the calculation of the standard QCD mechanism in the 0th flavor framework. The cross - sections will be estimated using as input the asymmetric intrinsic charm distributions provided by the CT18FC PDF grids \cite{Guzzi:2022rca}, corresponding to the meson-baryon model (MBM) for the intrinsic charm, as well as the grids from NNPDF40 PDF \cite{NNPDF:2023tyk}, obtained without any model assumptions, but which assume the charm and anticharm distributions as distinct degrees of freedom. In our analysis, we will use the central (best) fit parametrizations provide by these groups.
In addition, we also take into consideration analytical forms of the intrinsic charm PDFs as provided by Hobbs, Londergan and Melnitchouk (HLM) in Ref. \cite{Hobbs:2013bia}, which allow us to investigate the impact of an enhanced intrinsic charm component. For completeness of our analysis, in the Appendix, we present the predictions for the rapidity and transverse momentum distributions derived considering symmetric charm distributions, also available in the  CT18FC parametrization, which are based on the BHPS model.
Distinctly from the studies performed in Refs.~\cite{Maciula:2021orz,Maciula:2022otw}, which have  applied the fragmentation procedure in the overall center-of-mass system (c.m.s), in the current analysis, the fragmentation will be included after the boost to the parton-parton c.m.s., as proposed in Ref. \cite{Szczurek:2020vjn}. In the Appendix, we discuss in more detail the dependence of our predictions on this procedure.


\subsection{Results for the differential distributions}

\begin{figure}[!h]
\begin{minipage}{0.45\textwidth}
  \centerline{\includegraphics[width=1.0\textwidth]{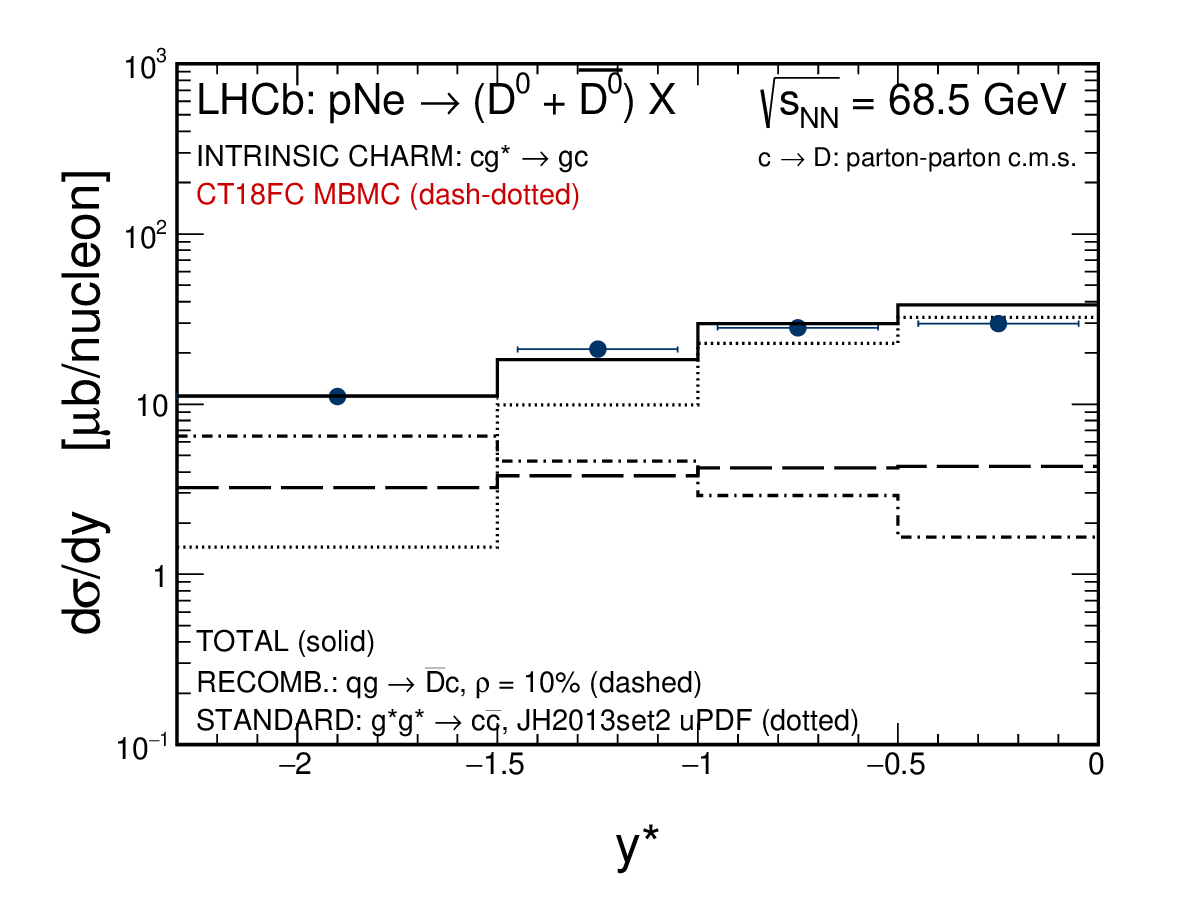}}
\end{minipage}
\begin{minipage}{0.45\textwidth}
  \centerline{\includegraphics[width=1.0\textwidth]{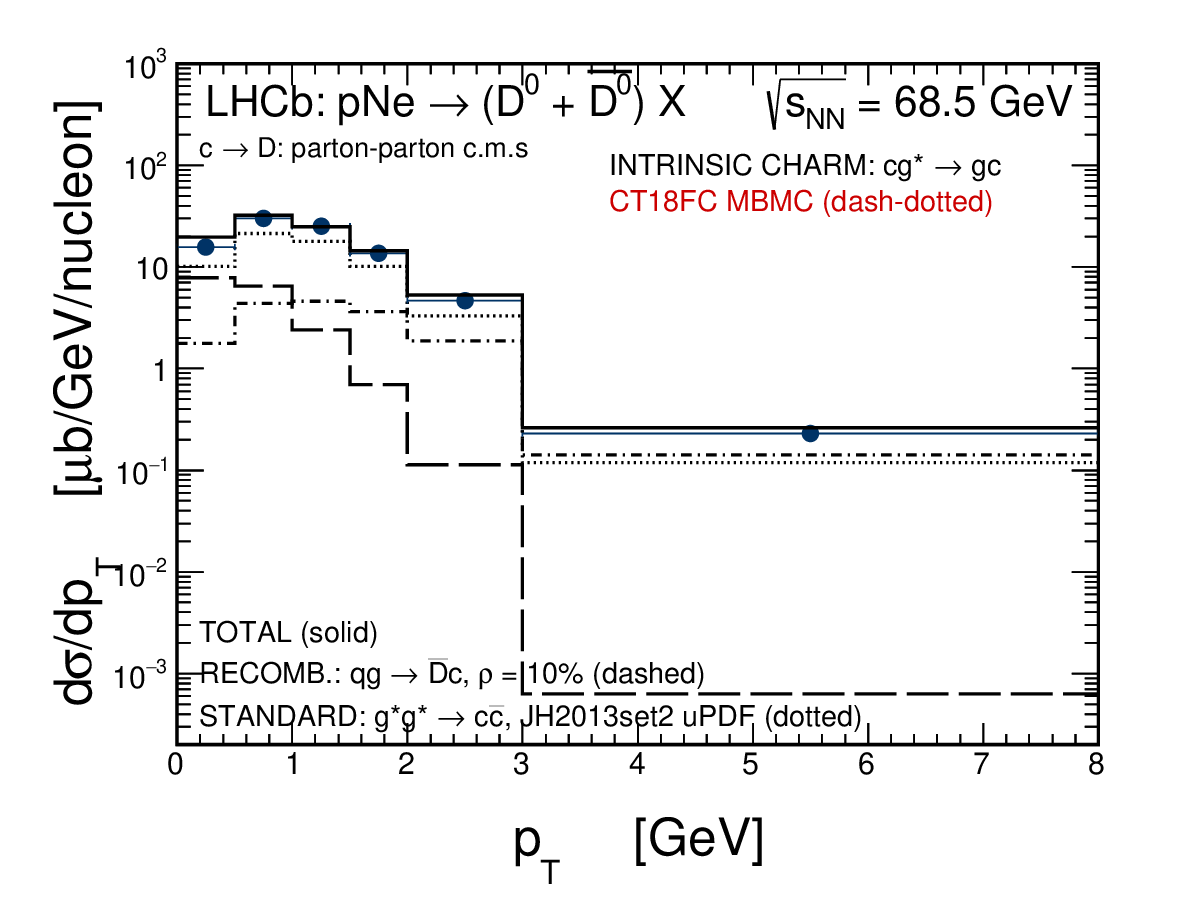}}
\end{minipage}
\begin{minipage}{0.45\textwidth}
  \centerline{\includegraphics[width=1.0\textwidth]{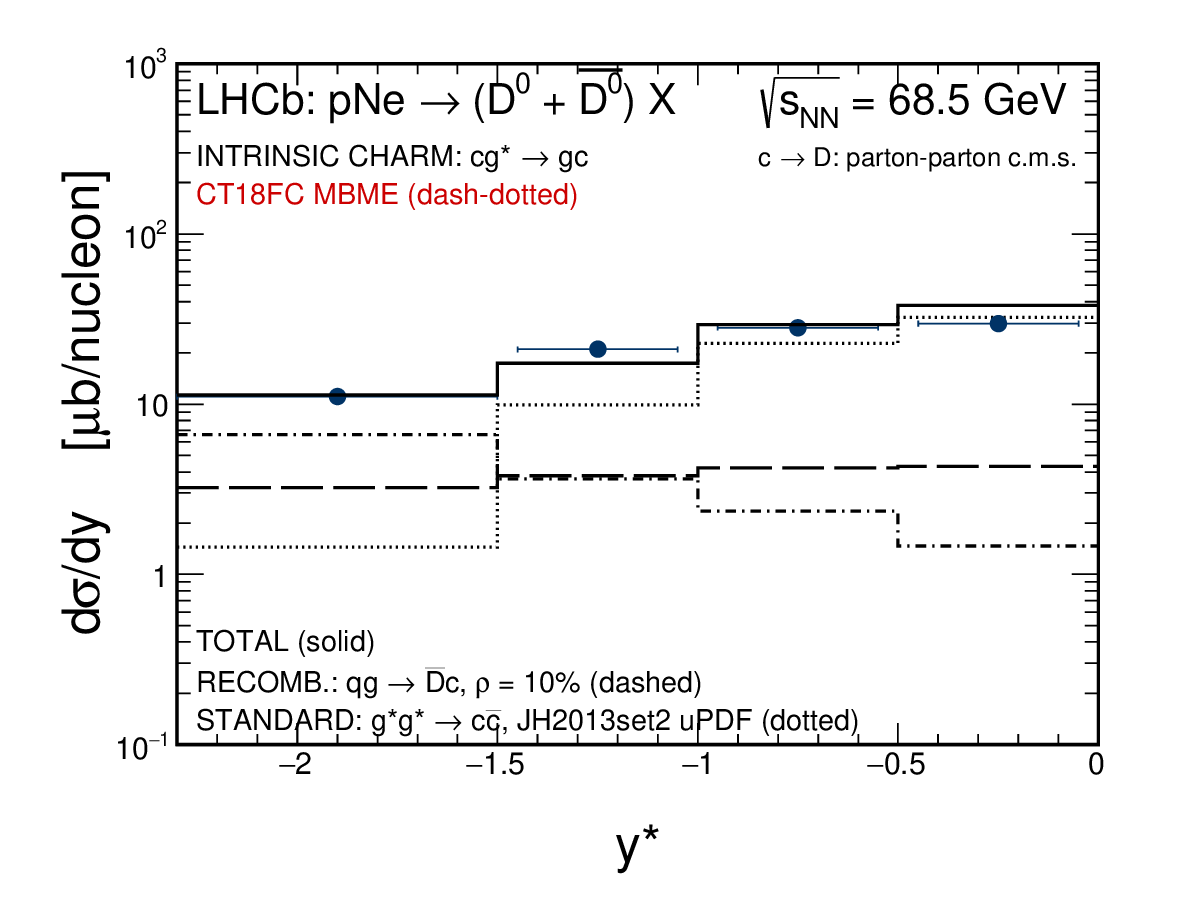}}
\end{minipage}
\begin{minipage}{0.45\textwidth}
  \centerline{\includegraphics[width=1.0\textwidth]{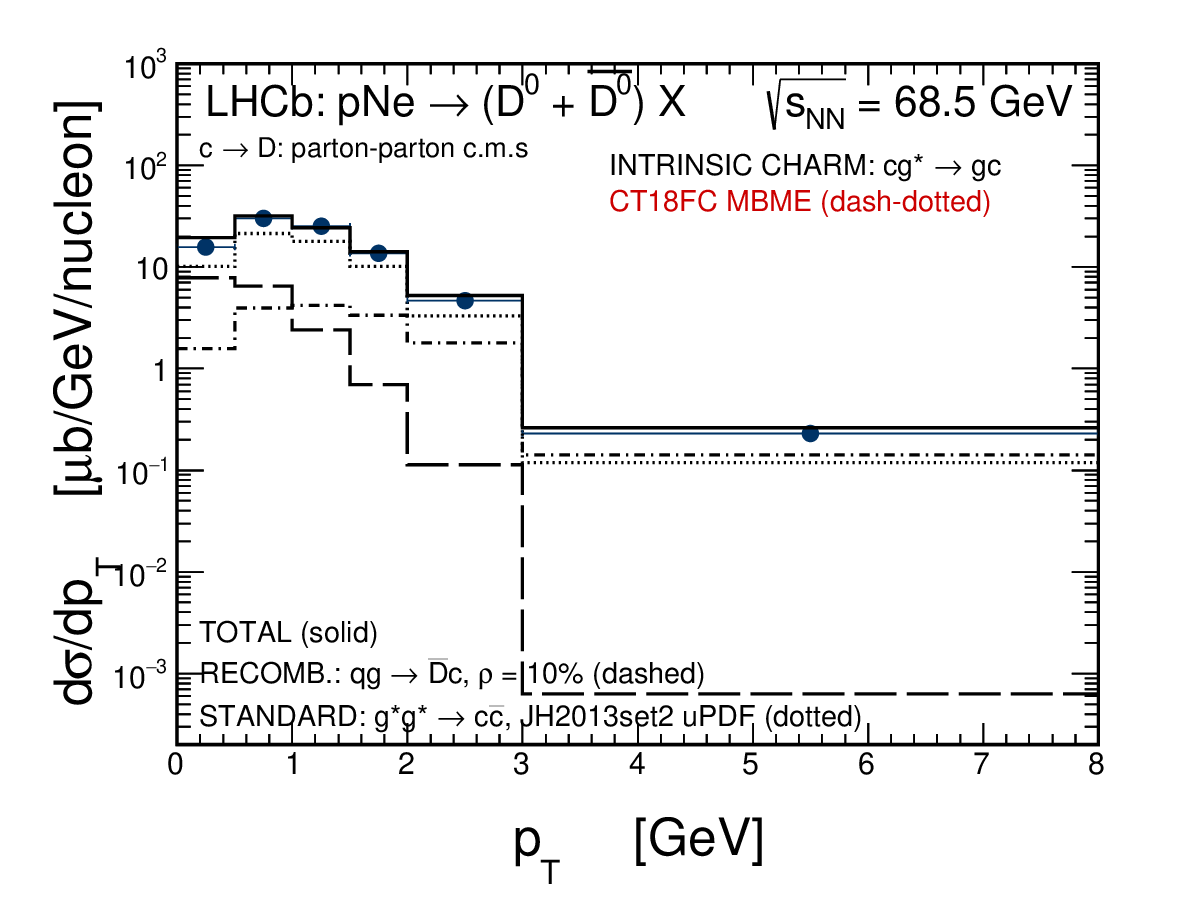}}
\end{minipage}
\begin{minipage}{0.45\textwidth}
  \centerline{\includegraphics[width=1.0\textwidth]{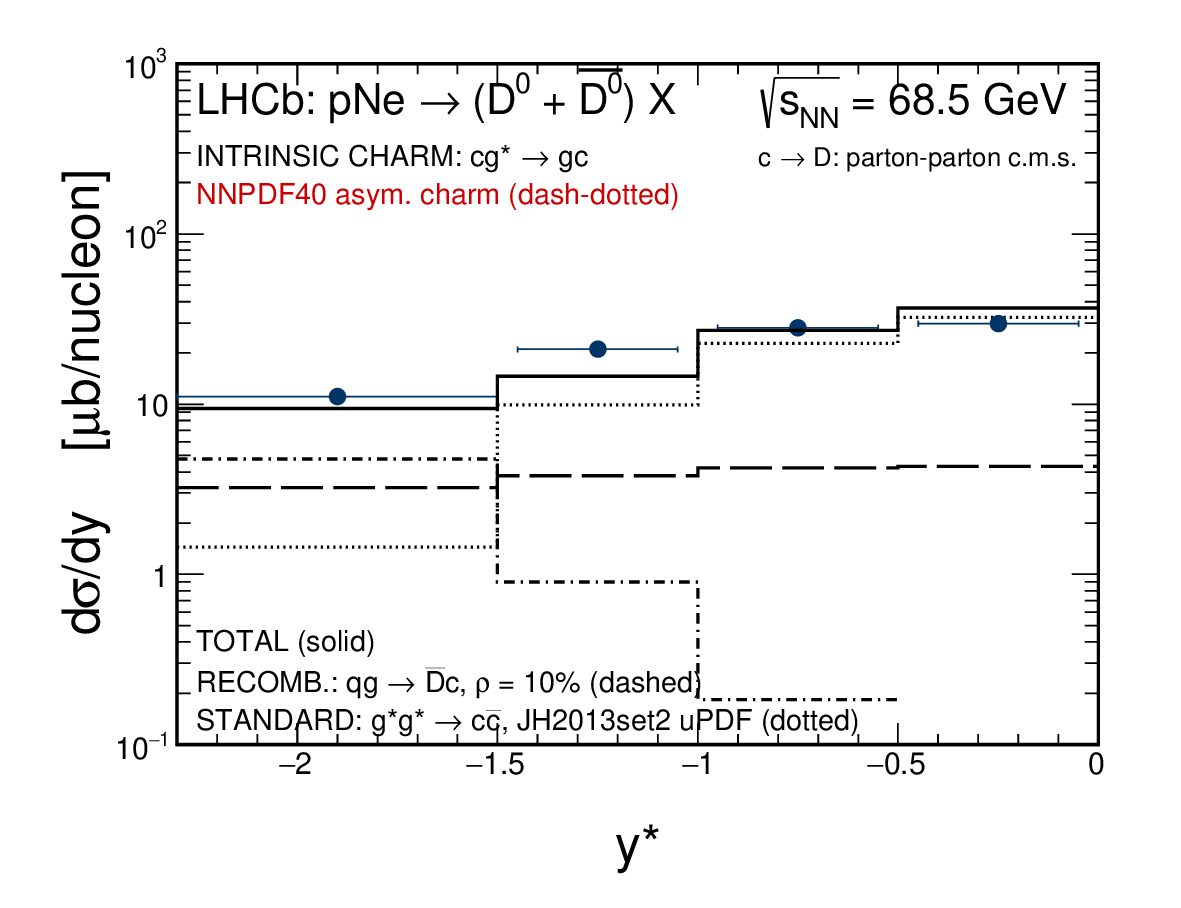}}
\end{minipage}
\begin{minipage}{0.45\textwidth}
  \centerline{\includegraphics[width=1.0\textwidth]{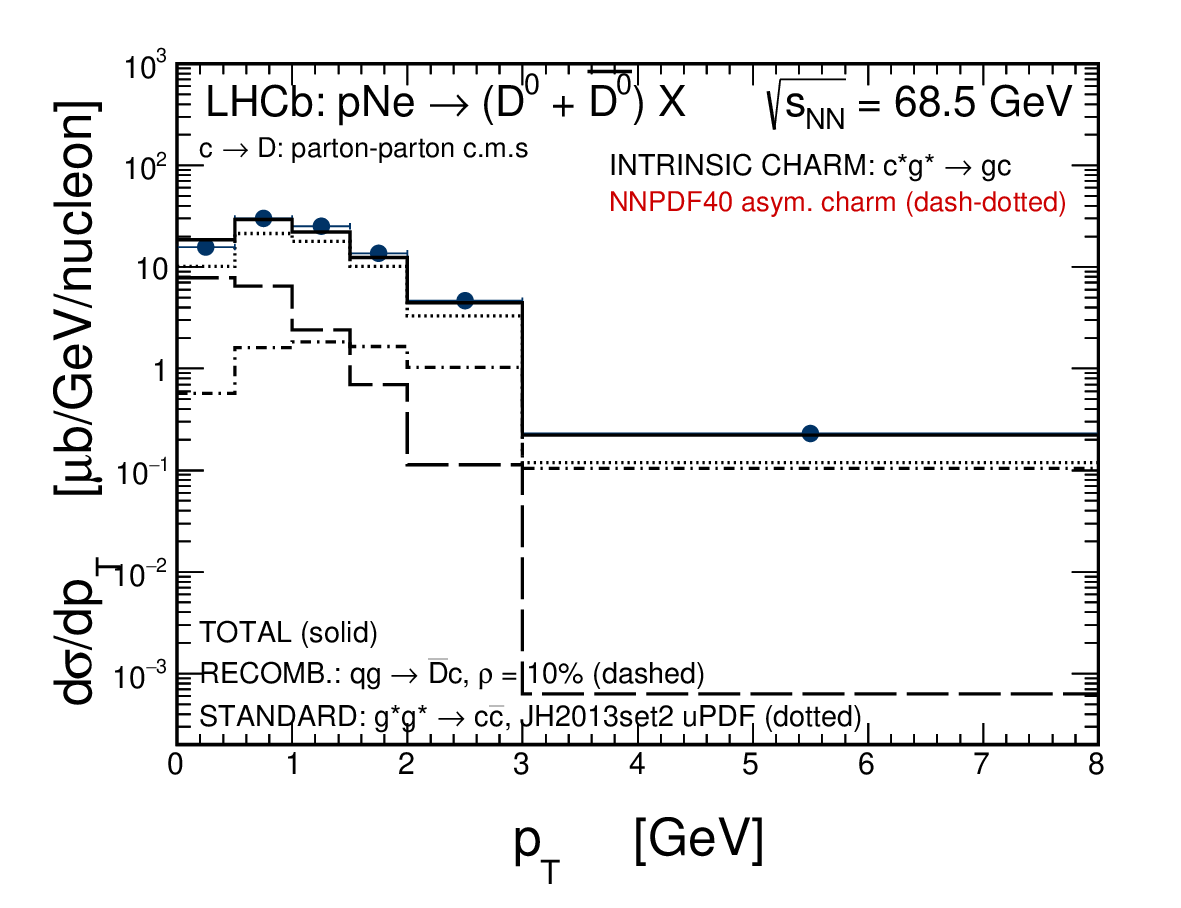}}
\end{minipage}
  \caption{\small The rapidity (left) and transverse momentum (right) distributions of $D^{0}$ meson (plus $\overline{D^{0}}$ antimeson) for $p+^{20}\!\mathrm{Ne}$ collisions at $\sqrt{s} = 68.5$ GeV together with the LHCb data \cite{LHCb:2022cul}. The three different contributions to charm meson production are shown separately, including the standard $g^*g^*\to c\bar c$ mechanism (dotted), the gluon - charm contribution (dot-dashed) and the recombination component (dashed). The solid histograms correspond to the sum of all considered mechanisms. Results derived using the CT18FC MBMC (upper panels), CT18FC MBME (middle panels) and NNPDF40 (lower panels) parametrizations are shown. }
\label{fig:dist_difPDFs}
\end{figure}

In Fig.~\ref{fig:dist_difPDFs} we present our predictions for the  rapidity (left panel) and transverse momentum (right panel) distributions of $D^{0}$ meson (plus $\overline{D^{0}}$ antimeson) in $p+^{20}\!\mathrm{Ne}$ collisions at $\sqrt{s} = 68.5$ GeV together with the LHCb data \cite{LHCb:2022cul}. The results associated with the three different contributions to charm meson production are shown separately, including the standard $g^*g^*\to c\bar c$ mechanism (dotted histograms), the gluon - charm contribution (dot-dashed histograms) and the recombination mechanism (dashed histograms). The solid histograms correspond to the sum of all considered mechanisms. Here, for the intrinsic charm distributions we use the CT18FC MBMC (upper panels), CT18FC MBME (middle panels) and NNPDF40 (lower panels) sets. One has that the inclusion of the gluon - charm mechanism improves the description of the experimental data. In particular, in the region of the most backward rapidities both the gluon - charm and the recombination contributions dominate over the standard mechanism. There, the gluon - charm contribution is about factor 2 and factor 5 larger than the recombination and the standard mechanism, respectively. Considering the transverse momentum spectra, we see that at low $p_{T}$'s it is dominated by the standard contribution, however, for the highest transverse momentum bin the contribution from the gluon - charm mechanism starts to play an important role. For large values of $p_{T}$, the standard contribution seems to underestimate the experimental data point, and the inclusion of the intrinsic charm component visibly improves the situation. The recombination contribution is found to be completely negligible at large meson transverse momenta\footnote{In fact, its contribution here might be enhanced by calculating higher order corrections to the recombination. However, at the moment, there is a lack of such formalism in the literature. We do not expect that the higher-order corrections would significantly change the overall picture here. Note, that the recombination contribution in the highest meson transverse momentum bin is more than two orders of magnitude smaller than the leading contributions.}.
Although the data indicates that the inclusion of the IC contribution is important, we also found that the quality of the description of the LHCb data for the three distinct models of asymmetric intrinsic charm is very similar. There are no big differences in the obtained normalization and shapes of the differential cross-sections for the inclusive single charm meson spectra between the CT18FC MBMC, CT18FC MBME and NNPDF40 parametrizations.

\begin{figure}[t]
\begin{minipage}{0.45\textwidth}
  \centerline{\includegraphics[width=1.0\textwidth]{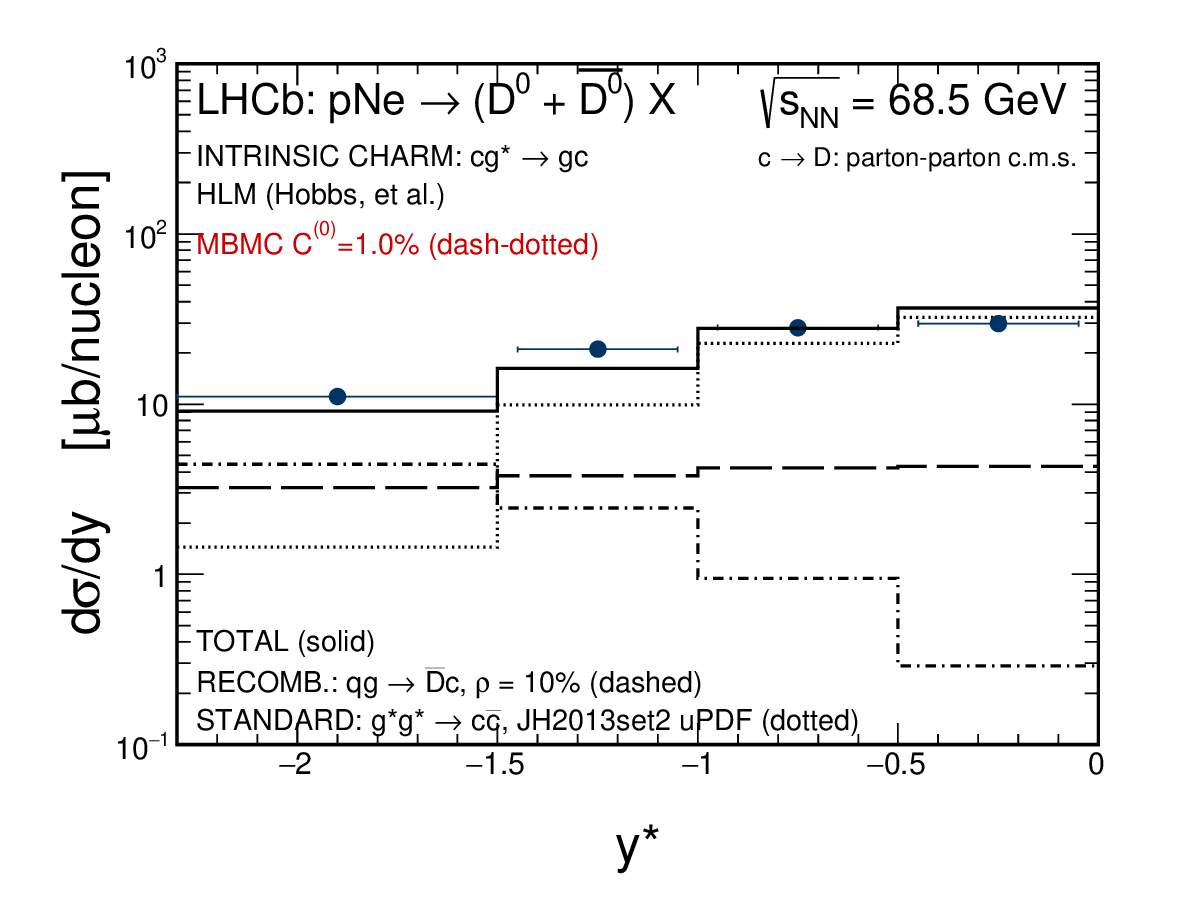}}
\end{minipage}
\begin{minipage}{0.45\textwidth}
  \centerline{\includegraphics[width=1.0\textwidth]{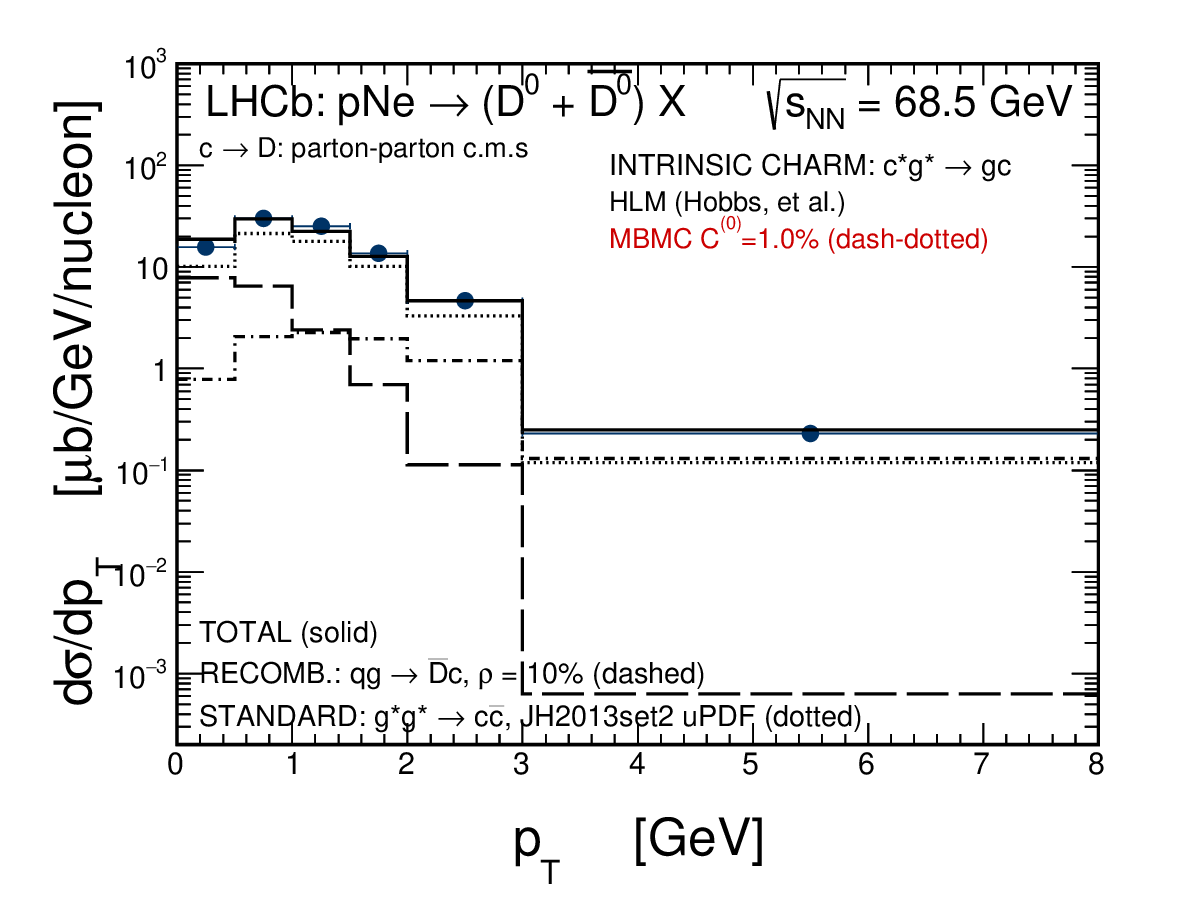}}
\end{minipage}
\begin{minipage}{0.45\textwidth}
  \centerline{\includegraphics[width=1.0\textwidth]{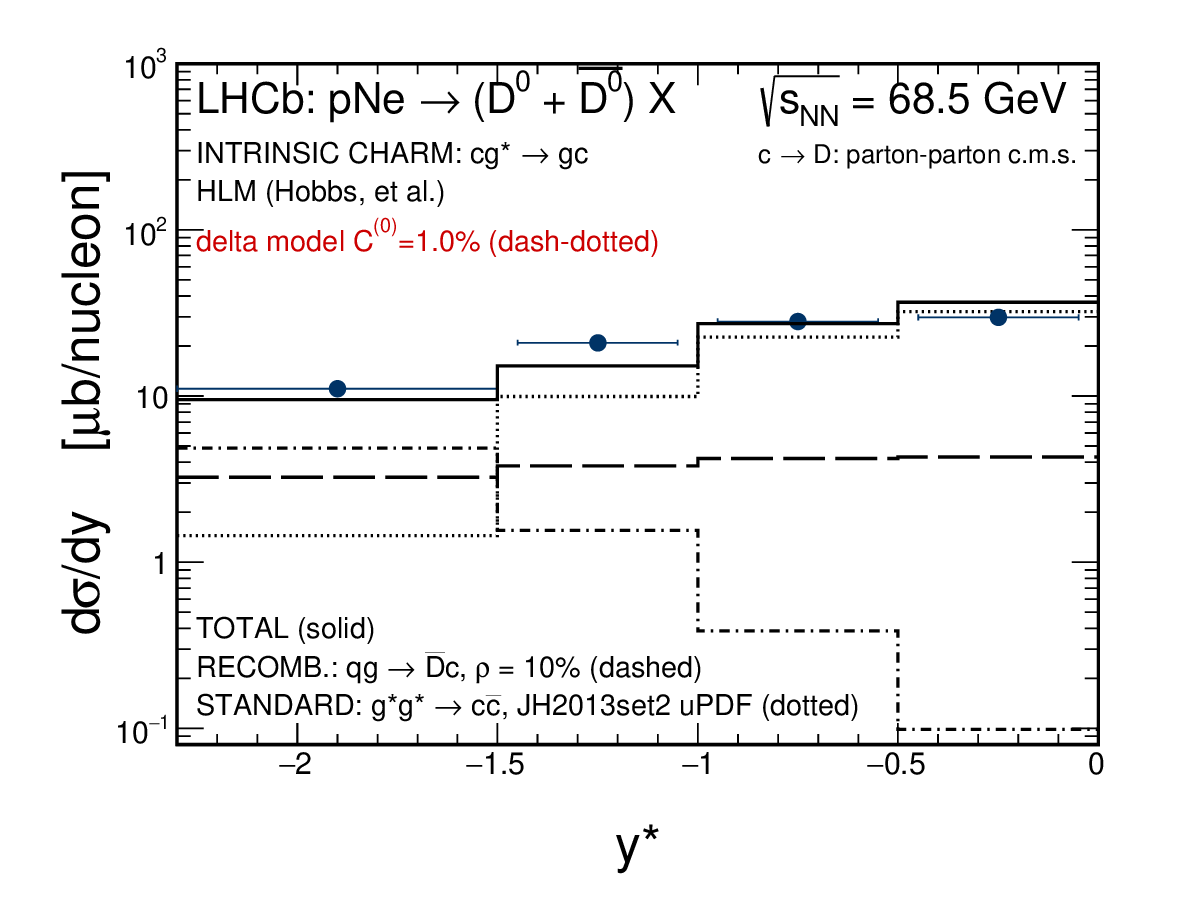}}
\end{minipage}
\begin{minipage}{0.45\textwidth}
  \centerline{\includegraphics[width=1.0\textwidth]{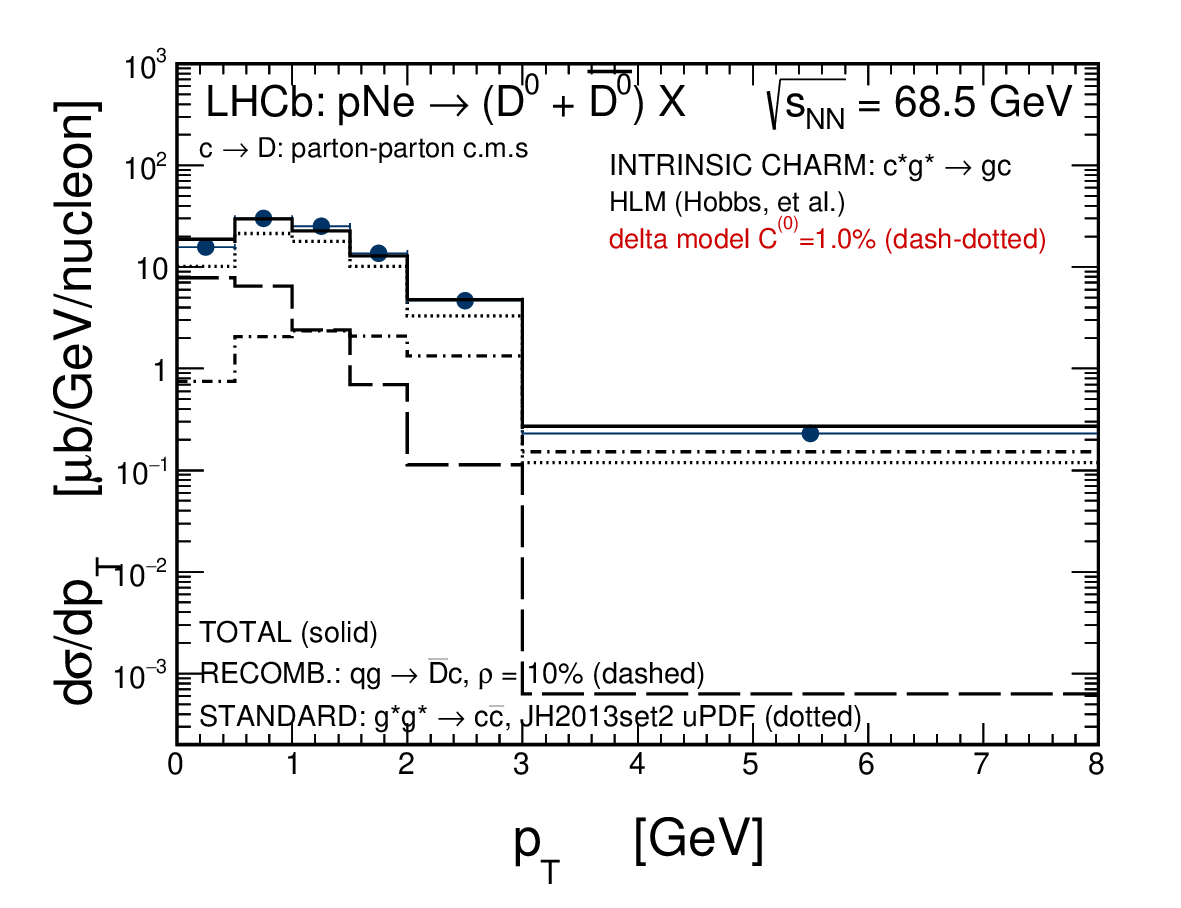}}
\end{minipage}
  \caption{
\small The rapidity (left) and transverse momentum (right) distributions of $D^{0}$ meson (plus $\overline{D^{0}}$ antimeson)
for $p+^{20}\!\mathrm{Ne}$ collisions at $\sqrt{s} = 68.5$ GeV together with the LHCb data \cite{LHCb:2022cul}. Here three different contributions to charm meson production are shown separately, including the standard $g^*g^*\to c\bar c$ mechanism (dotted), the gluon - charm contribution (dot-dashed) and the recombination component (dashed). The solid histograms correspond to the sum of all considered mechanisms. Here,  we use the  HLM parametrizations ~\cite{Hobbs:2013bia}  for the intrinsic charm PDFs within the MBMC (upper panels) and the $\delta$-function
(lower panels) models. Details are specified in the figure.
}
\label{fig:8}
\end{figure}

In order to verify whether the current data is able to discriminate possible descriptions of the intrinsic charm, we also have investigated the implication of alternative models. In particular,  we have also taken into consideration the intrinsic charm parametrizations presented in Ref.~\cite{Hobbs:2013bia} by Hobbs, Londergan and Melnitchouk (HLM).
There, the authors provided simple three-parameter fits to the intrinsic charm quark distributions within
the MBM, computed using several different models for the $c$ and $\bar c$ distributions in the
charmed mesons and baryons, including the confining model, effective mass model, and the
$\delta$-function model.    
In Fig.~\ref{fig:8}, we show our results obtained with the HLM parametrizations for the intrinsic charm PDFs within the MBMC (upper  panels) and the $\delta$-function (lower panels) models, respectively. These parametrizations allow some freedom in varying their normalization. Here we have normalized the PDFs taking the normalization constant $C^{(0)}=1.0\%$. Within this choice, we obtain very similar results as previously in the case of the intrinsic charm distributions from the CT18FC PDFs. Again, the quality of the description of the LHCb data for both the rapidity and meson transverse momentum distributions is very good.

Summarizing, we do not find any significant differences in our predictions for the rapidity and transverse momentum distributions for charm mesons arising from the use of different up-to-date intrinsic charm PDFs from the literature.

\subsection{Production asymmetry}

An alternative to probe possible asymmetries in the charm and anticharm distributions is the comparison between the cross - sections for the production of $D^0$ and ${\overline D}^0$ mesons. In the last years, the LHCb Collaboration has released data for the production asymmetry in $p+^{20}\!\mathrm{Ne}$ collisions at $\sqrt{s} = 68.5$ GeV   \cite{LHCb:2022cul}, defined by 
\begin{equation}
A_{p} = \frac{d\sigma^{D^0}\!\!/d\xi - d\sigma^{\overline{D}^0}\!\!/d\xi}
                   {d\sigma^{D^0}\!\!/d\xi + d\sigma^{\overline{D}^0}\!\!/d\xi}
\; ,
\label{asymmetry}
\end{equation}
where $\xi$ represents a single variable ($y$ or $p_t$) or even a pair of
variables ($(y, p_t)$). The experimental data indicates a sizeable asymmetry for negative rapidities and large transverse momentum, and the description of these results in the full kinematical range is still a theoretical challenge, since initial and final states can contribute to generate the observed asymmetry. In a previous study \cite{Maciula:2022otw}, it was demonstrated that an asymmetry can be produced within the recombination mechanism, but the predictions are not able to describe the data for large transverse momentum. Here, we will investigate if an asymmetric charm distribution, not considered in Ref. \cite{Maciula:2022otw}, can improve the description of the data and be used to discriminate between the distinct models for the IC component.

\begin{figure}[!h]
\begin{minipage}{0.45\textwidth}
  \centerline{\includegraphics[width=1.0\textwidth]{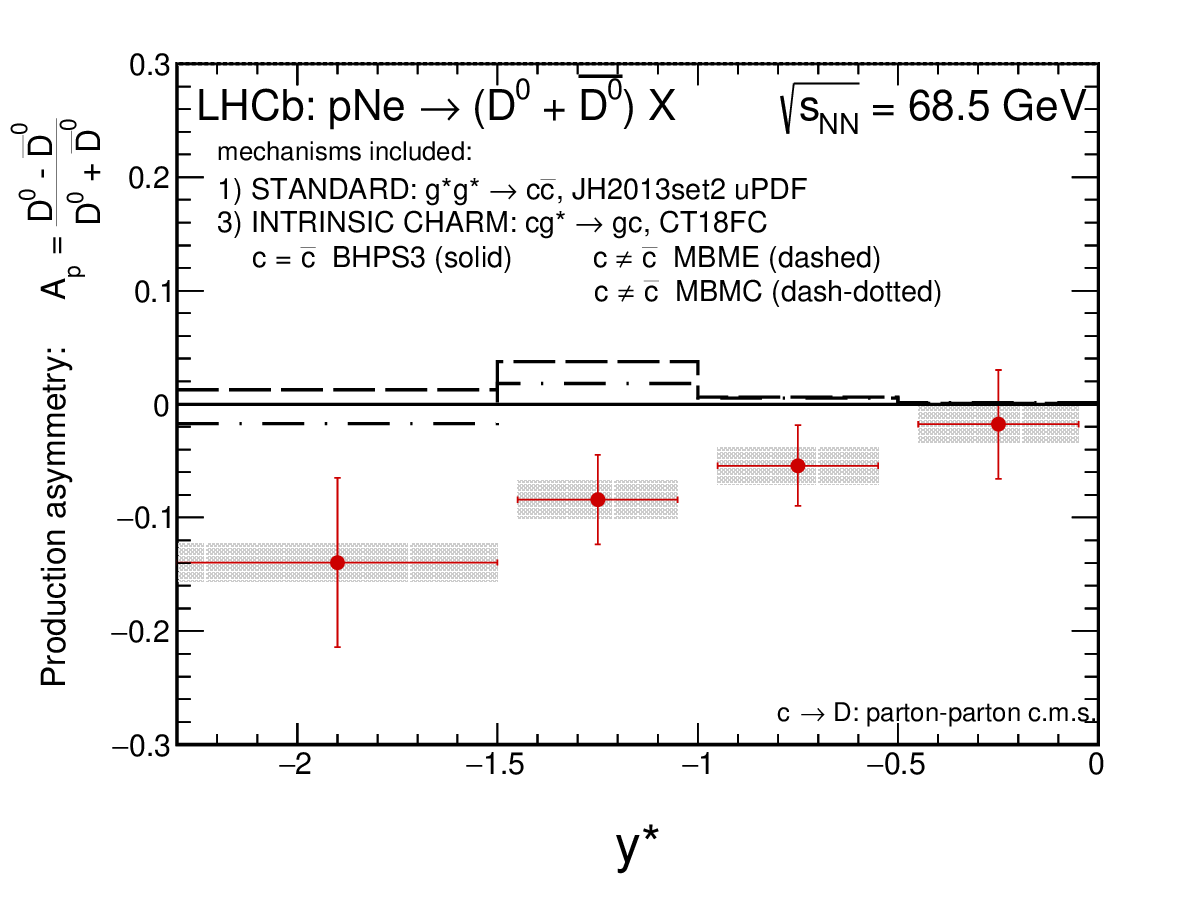}}
\end{minipage}
\begin{minipage}{0.45\textwidth}
  \centerline{\includegraphics[width=1.0\textwidth]{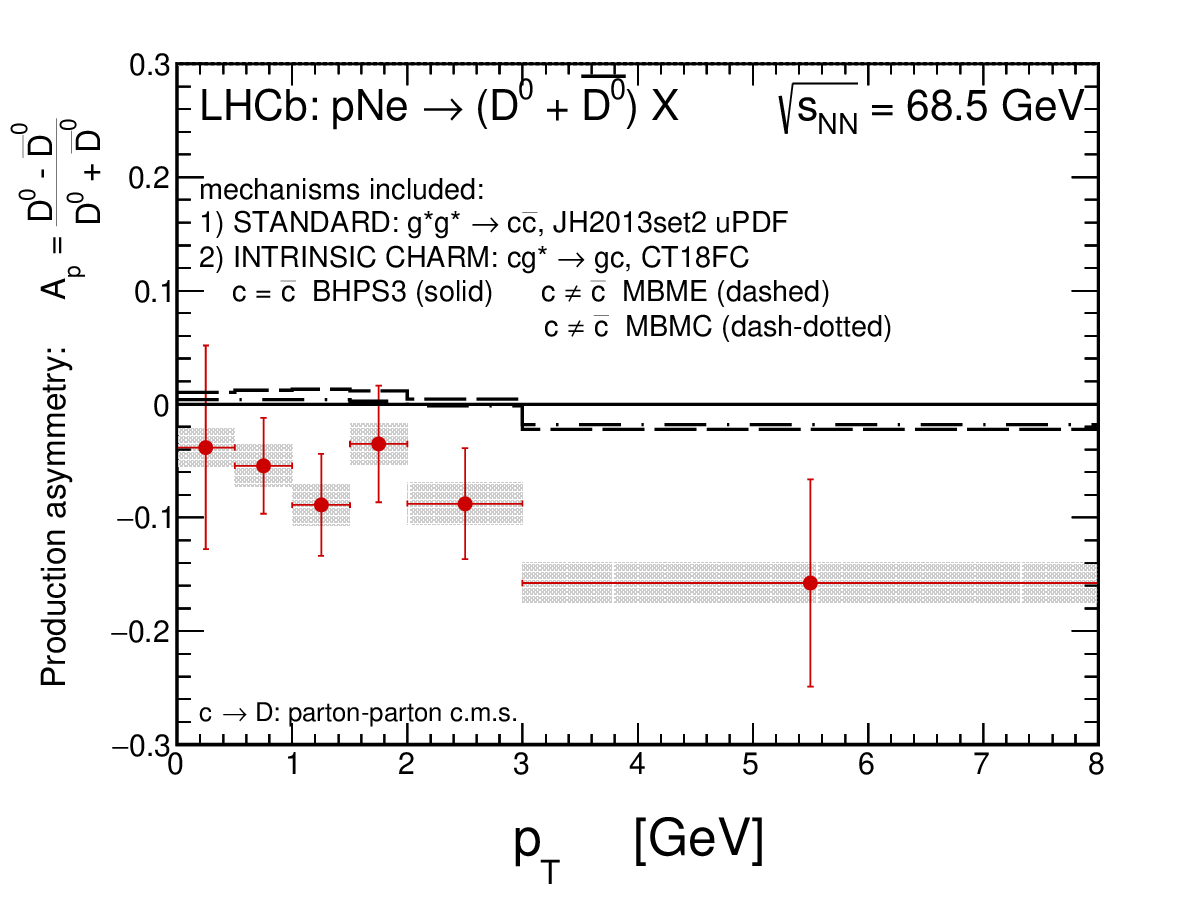}}
\end{minipage}
\begin{minipage}{0.45\textwidth}
  \centerline{\includegraphics[width=1.0\textwidth]{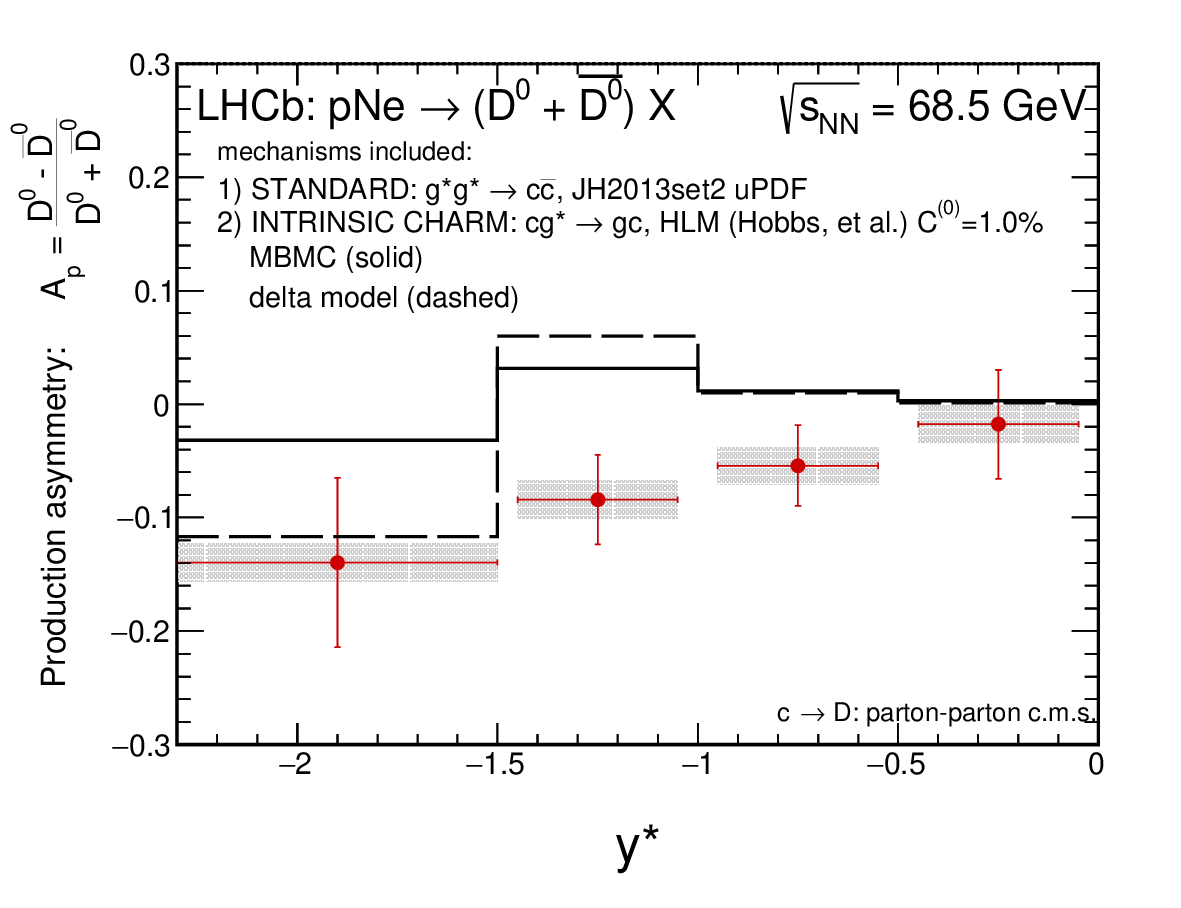}}
\end{minipage}
\begin{minipage}{0.45\textwidth}
  \centerline{\includegraphics[width=1.0\textwidth]{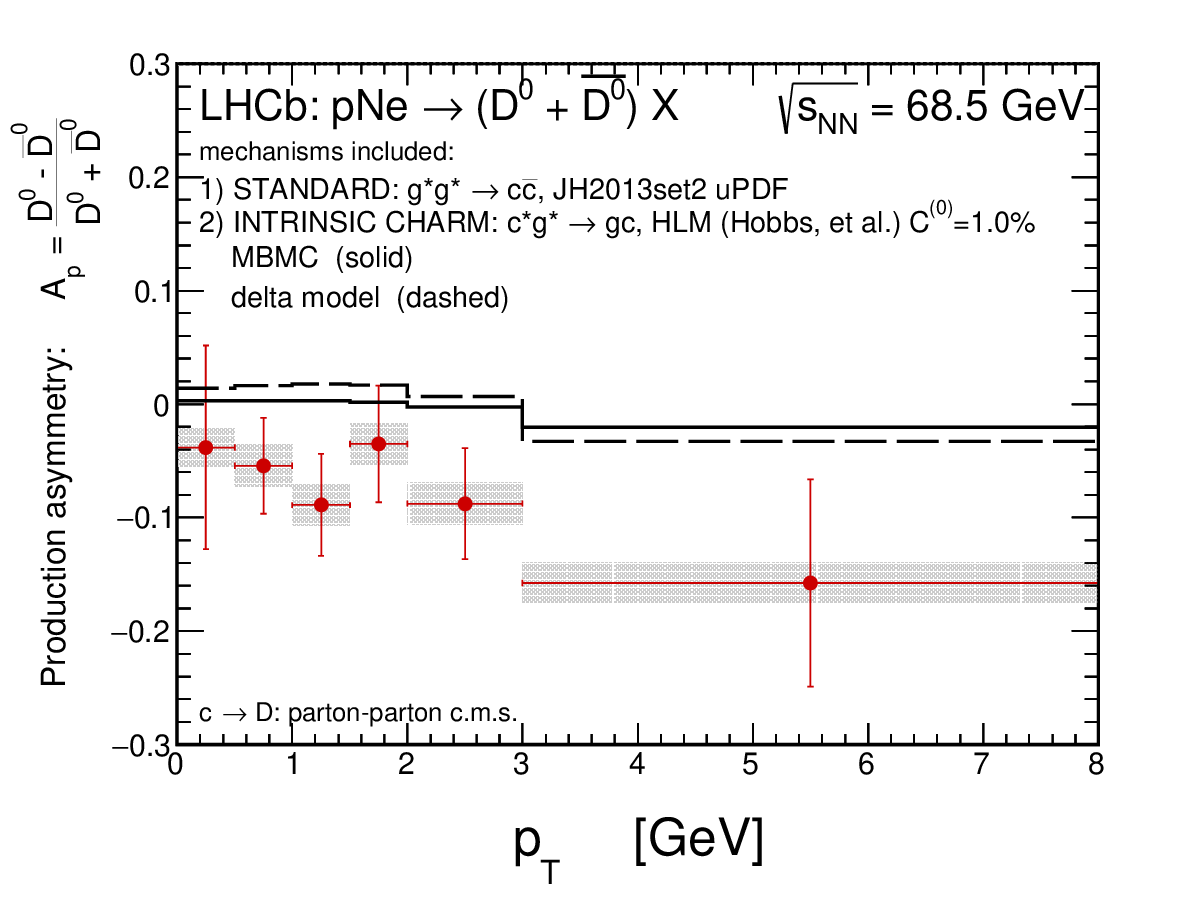}}
\end{minipage}
\begin{minipage}{0.45\textwidth}
  \centerline{\includegraphics[width=1.0\textwidth]{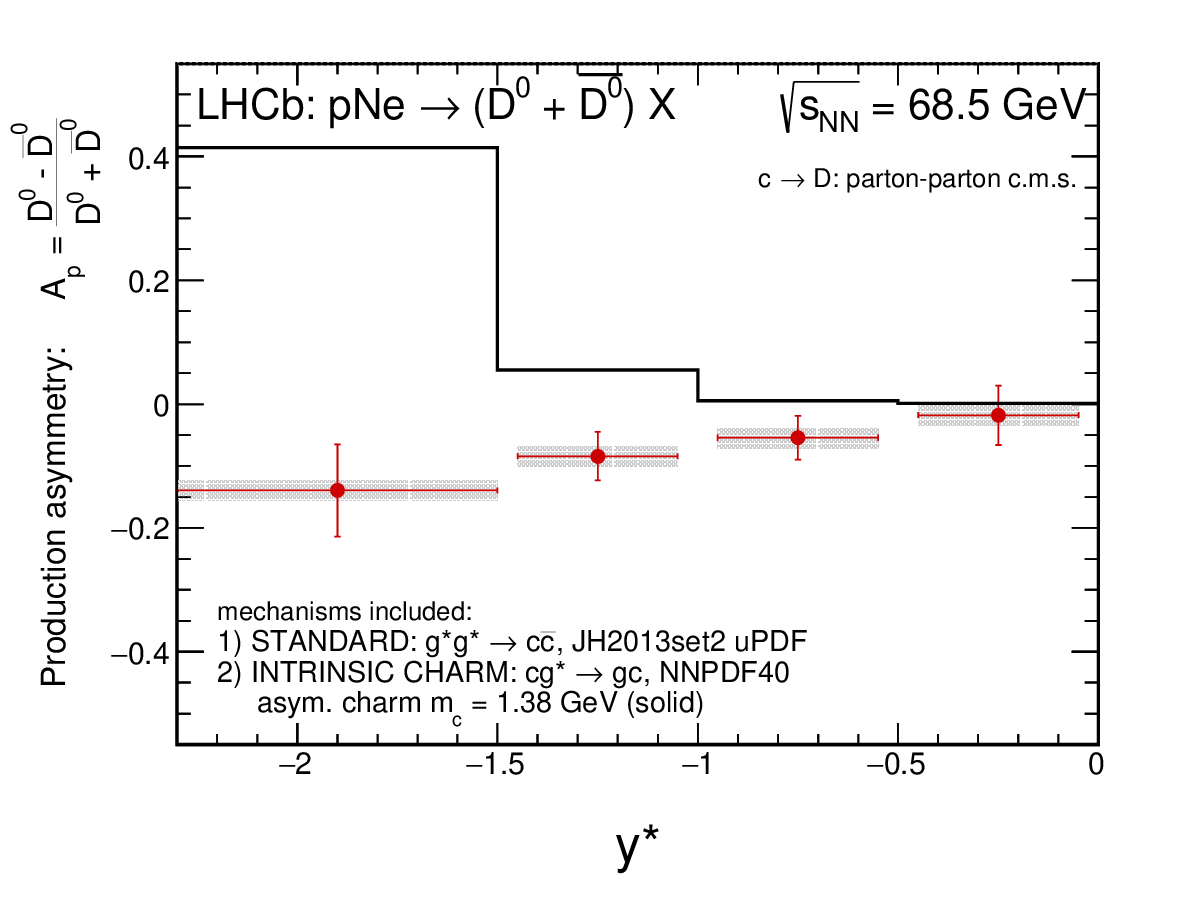}}
\end{minipage}
\begin{minipage}{0.45\textwidth}
  \centerline{\includegraphics[width=1.0\textwidth]{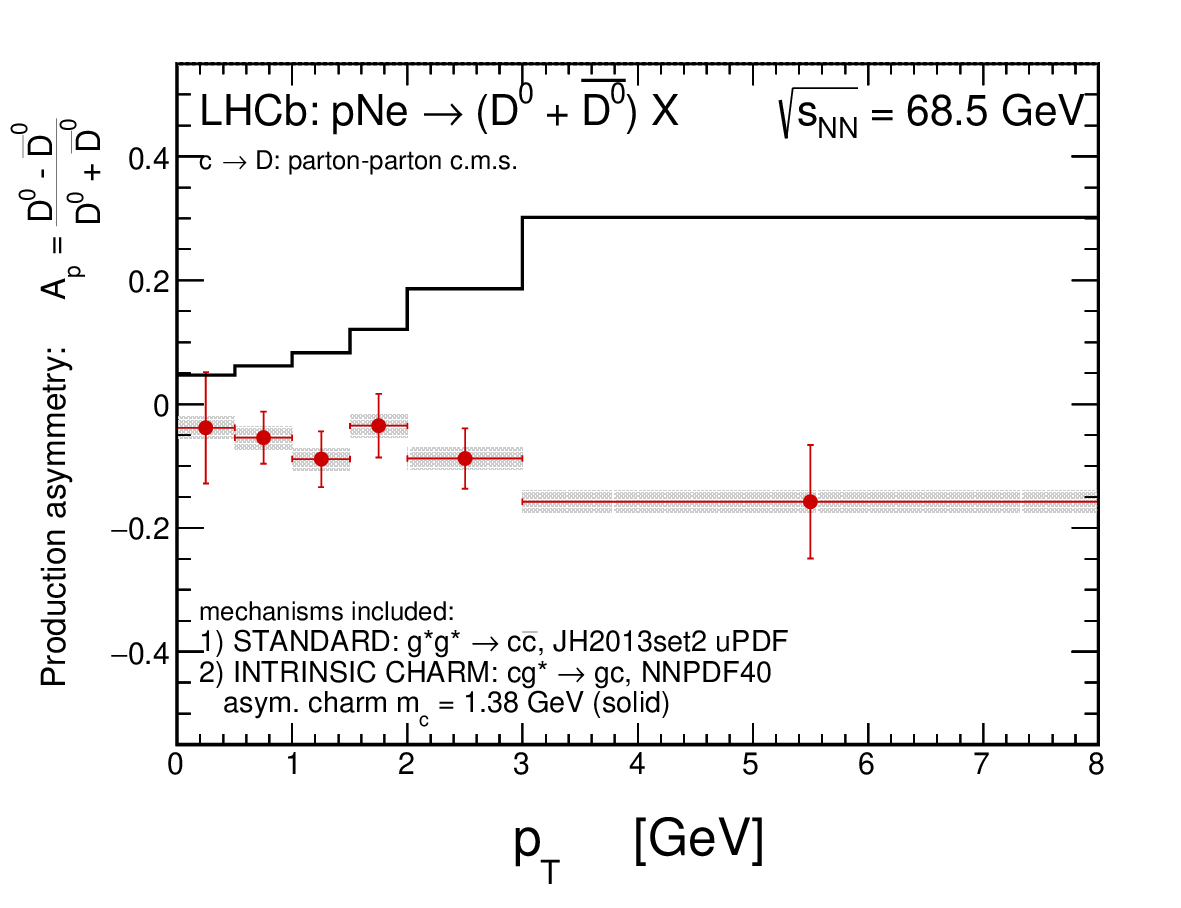}}
\end{minipage}
  \caption{
\small The production asymmetry $A_{p}$ for $D^{0}$-meson and $\overline{D}^{0}$-antimeson as a function of rapidity (left) and transverse momentum (right) for $p+^{20}\!\mathrm{Ne}$ collisions at $\sqrt{s} = 68.5$ GeV together with the LHCb data \cite{LHCb:2022cul}. Results derived assuming the standard $g^*g^*\to c\bar c$ and gluon - charm mechanisms and different parametrizations for the intrinsic charm distributions:  CT18FC BHPS3 and MBM  (upper panels), HLM  (middle panels) and NNPDF40 (lower panels). Details are specified in the figure.
}
\label{fig:asym_gc}
\end{figure}

Initially, in Fig. \ref{fig:asym_gc}, we present our predictions for the production asymmetry, $A_{p}$, derived taking into account only of the gluon - gluon and  gluon - charm mechanisms and considering distinct models for the intrinsic charm and anticharm distributions. Distinctly from the gluon - gluon mechanism, which does not generate asymmetries, the $gc \rightarrow gc$ and $g\bar{c} \rightarrow g\bar{c}$ reactions can generate an asymmetry if $c(x) \neq \bar{c}(x)$, with the predictions being directly dependent on the model assumed for the asymmetric charm distributions. Such an expectation is observed in the results presented in  Fig. \ref{fig:asym_gc}. In particular, in the upper panels, we present the predictions derived using the CT18FC MBMC and  CT18FC MBME parametrizations. For comparison, the results obtained using as input the CT18FC BHPS parametrization, which assumes  $c(x) = \bar{c}(x)$ in the initial condition, are also shown.
The solid histograms correspond to the symmetric intrinsic charm encoded in the BHPS3 PDFs that naturally lead to the symmetric production of $D$-meson and $\bar D$-antimeson, i.e. $A_{p} = 0$. The dashed and dash-dotted histograms correspond to the MBME and MBMC intrinsic charm PDFs, respectively.
We observe that both the MBM distributions result in a non-zero production asymmetry, however, the corresponding theoretical values are much smaller than those measured by the LHCb. 

The results for the  HLM intrinsic charm distributions are presented in the middle panels of Fig. \ref{fig:asym_gc}. In this case, the delta function model produces larger asymmetry than the MBMC model in the most backward rapidity bin and leads to the value compatible with the data. However, for the asymmetry as a function of the meson transverse momenta the predicted values underestimate the LHCb data points as in the case of the CT18FC distributions. The observed discrepancy is growing together with the meson $p_{T}$. In the MBM the peak in the charm distribution in a hadron is related to the fraction of the hadron mass carried by the charm quark, therefore the resulting distribution of $\bar c$ in the $\bar D$ meson is typically harder than that for the $c$ in the $\Lambda_c$. Thus, within our model, the MBM intrinsic charm PDFs result in more  $\bar c$-antiquarks (and $\overline{D^{0}}$-mesons) than $c$-quarks (and $D^{0}$-mesons) at large meson transverse momenta. However, the corresponding effect is by far too small.

 Finally, the results associated with the NNPDF40 parametrization for intrinsic charm distribution are presented in the lower panels of Fig. \ref{fig:asym_gc}. In this case, we get much larger asymmetries than in the case of the MBM PDFs, however, their sign is positive across the whole considered spectra, in contrast with the LHCb data. Within the NNPDF40 PDF we get significantly more charm quarks than charm antiquarks, and in consequence more $D^{0}$-mesons than $\overline{D^{0}}$-antimesons, especially at large meson transverse momenta and for backward rapidities.


\begin{figure}[!h]
\begin{minipage}{0.45\textwidth}
  \centerline{\includegraphics[width=1.0\textwidth]{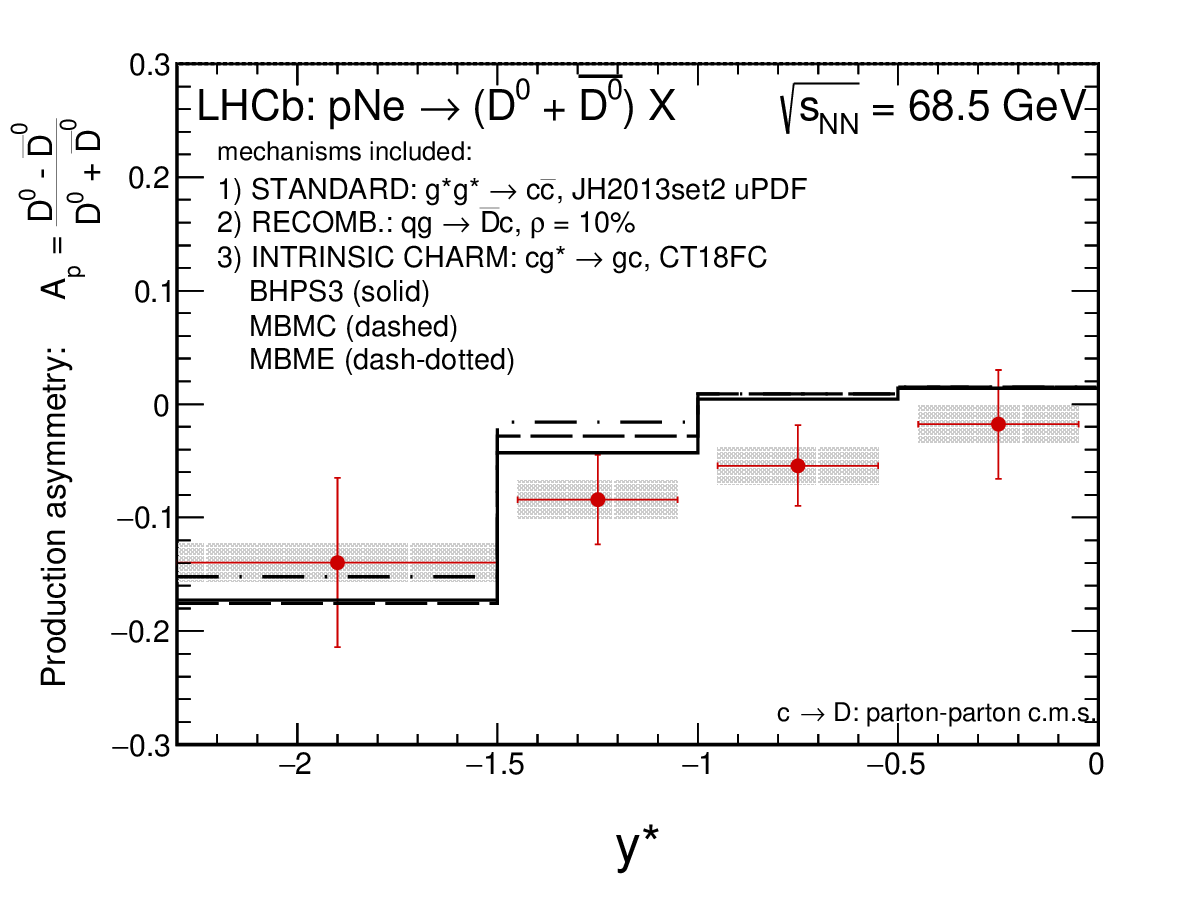}}
\end{minipage}
\begin{minipage}{0.45\textwidth}
  \centerline{\includegraphics[width=1.0\textwidth]{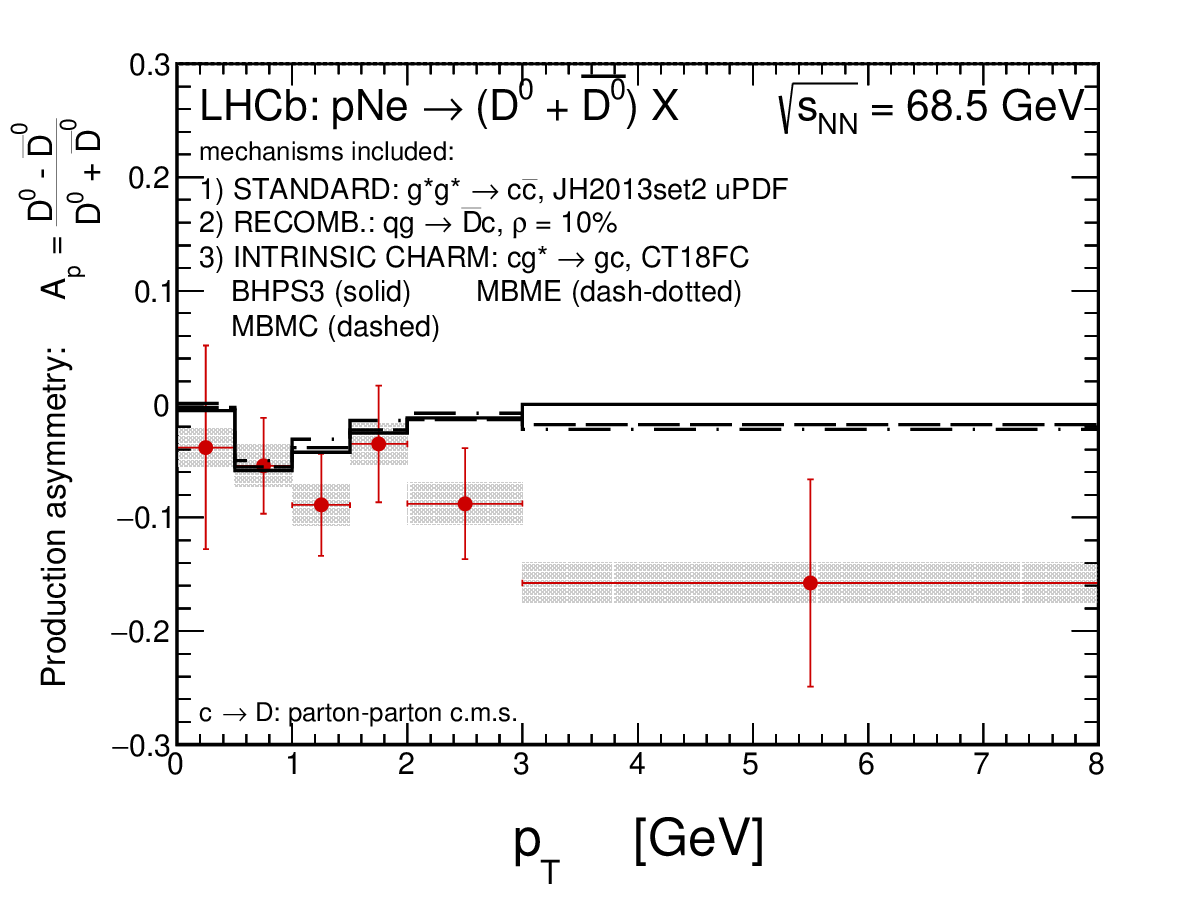}}
\end{minipage}
\begin{minipage}{0.45\textwidth}
  \centerline{\includegraphics[width=1.0\textwidth]{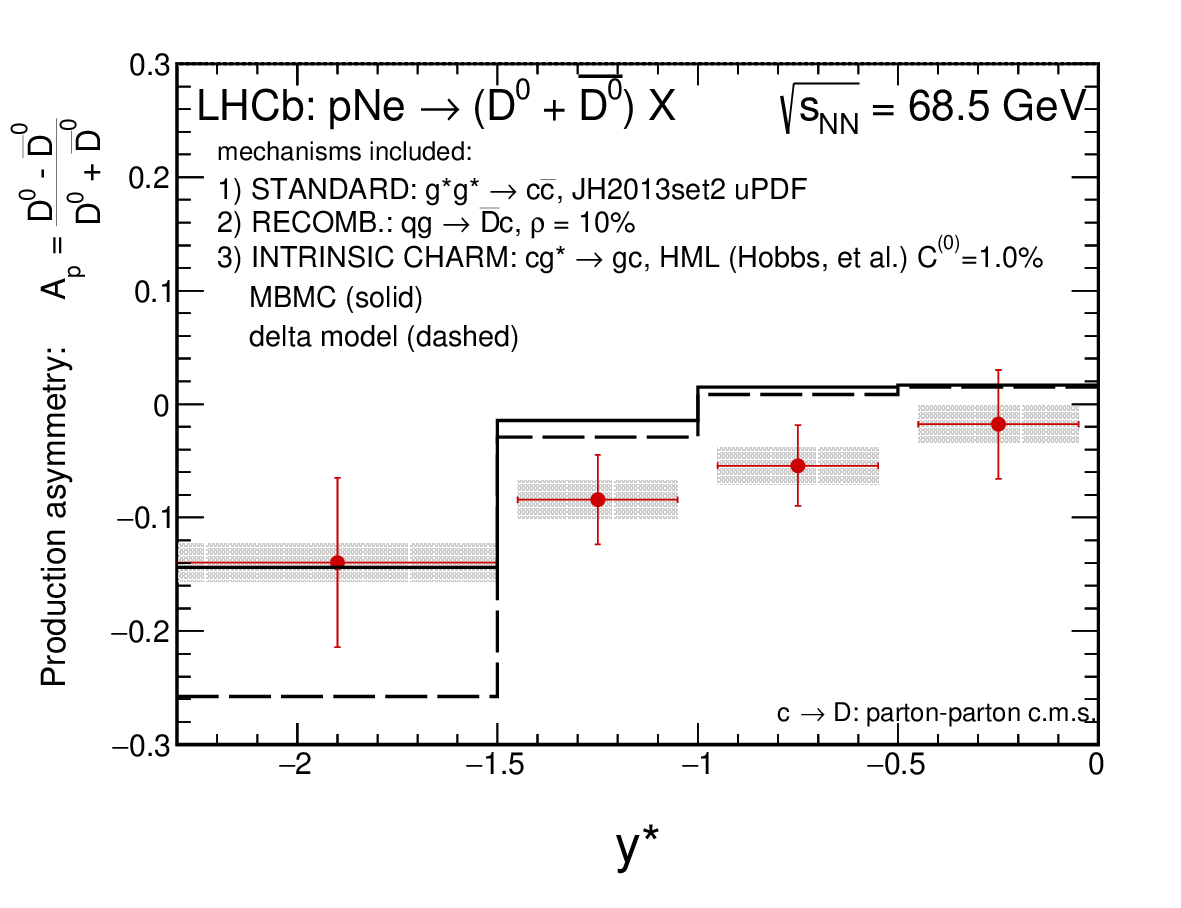}}
\end{minipage}
\begin{minipage}{0.45\textwidth}
  \centerline{\includegraphics[width=1.0\textwidth]{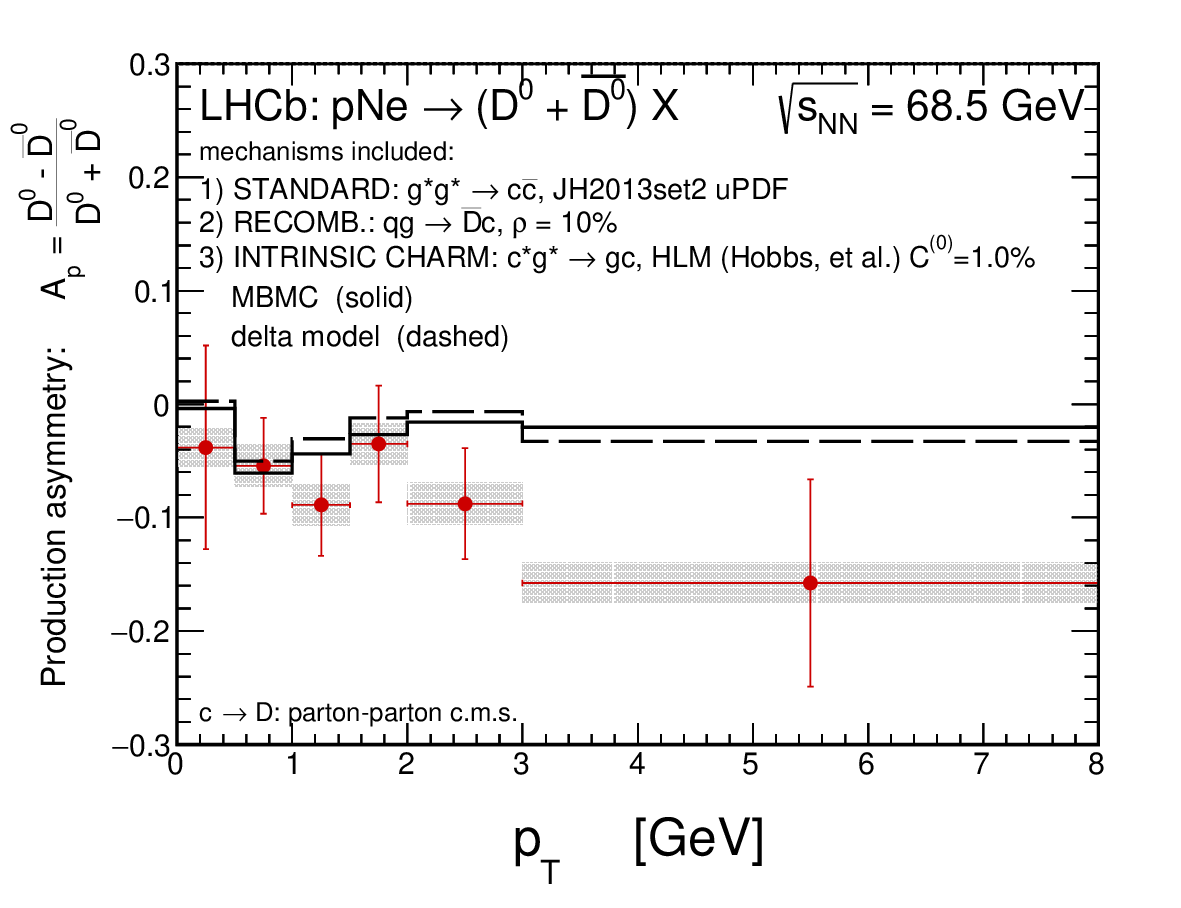}}
\end{minipage}
\begin{minipage}{0.45\textwidth}
  \centerline{\includegraphics[width=1.0\textwidth]{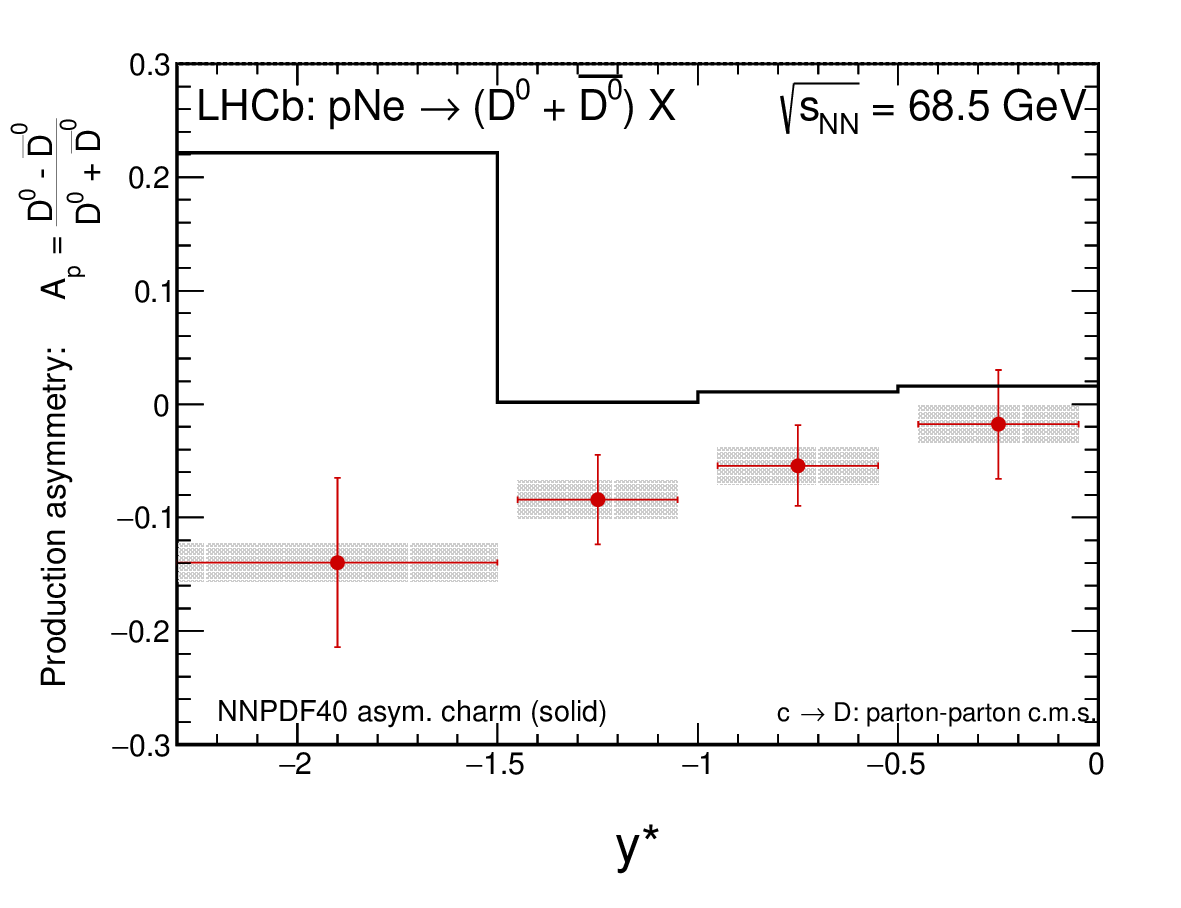}}
\end{minipage}
\begin{minipage}{0.45\textwidth}
  \centerline{\includegraphics[width=1.0\textwidth]{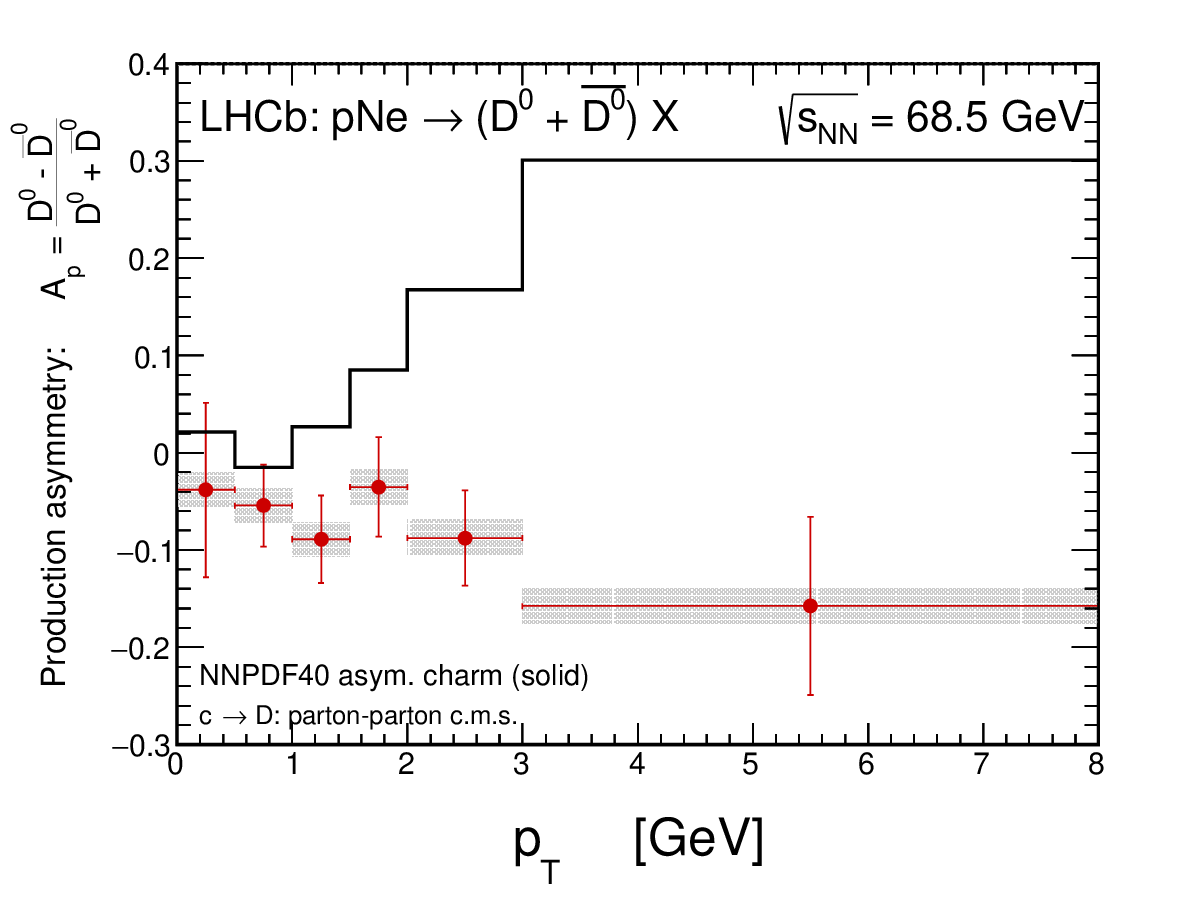}}
\end{minipage}
  \caption{
\small The production asymmetry $A_{p}$ for $D^{0}$-meson and $\overline{D}^{0}$-antimeson as a function of rapidity (left) and transverse momentum (right) for $p+^{20}\!\mathrm{Ne}$ collisions at $\sqrt{s} = 68.5$ GeV together with the LHCb data \cite{LHCb:2022cul}. Results derived assuming the standard $g^*g^*\to c\bar c$, gluon - charm and recombination mechanisms, and different parametrizations for the intrinsic charm distributions:  CT18FC MBM  (upper panels), HLM  (middle panels) and NNPDF40 (lower panels). Details are specified in the figure.
}
\label{fig:asym_sum}
\end{figure}

The results presented in Fig. \ref{fig:asym_gc} indicate that if only initial effects are taken into account, the current intrinsic charm models are not able to describe the current data for the production asymmetry. As pointed  out above, these data are also not described if only the recombination process is considered \cite{Maciula:2022otw}. In what follows, we will investigate if the sum of these mechanisms can improve the description. The associated results are presented in Fig. 
\ref{fig:asym_sum} for the distinct models of  intrinsic charm PDFs. One has that the data are better described when both mechanisms are summed and the PDFs are described by the CT18FC and HLM models (upper and middle panels). However, the description of the data for large transverse momentum is still a challenge. Surely, more data in this kinematical region is important in order to constrain the IC and recombination mechanisms. Finally, if the NNPDF40 model is used as input in our calculations, the predictions largely disagree with the current data.




\section{Summary}
\label{sec:conc}
One of the main challenges of the Particle Physics is the description of the proton structure. In recent years, several studies have indicated that in addition to the light up and down valence quarks, an intrinsic charm valence distribution can also be present in the proton wave function. The probe of existence of this intrinsic component and the determination of its properties are important aspects that have motivated several theoretical and experimental investigations. In this paper we have considered the possibility that the charm and anticharm distributions for a given value of the Bjorken - $x$ variable  are distinct and investigate the impact of this asymmetry on  $D^0$ and ${\bar D}^0$ meson production in fixed - target $p + ^{20}Ne$ collisions at the LHC. We have estimated 
the rapidity and transverse momentum distributions considering the more recent parametrizations for an asymmetric charm distribution. Our results indicated that the inclusion of an intrinsic charm component improves the description of the current LHCb data for these distributions, but is not able to discriminate between the distinct models for the description of the IC. Moreover, we also have estimated the production asymmetry, which is expected to be more sensitive to asymmetries on the initial state, without and with the inclusion of the recombination process. For the first case, we have obtained that the  intrinsic charm PDFs based on the meson cloud model generate a small amount of asymmetry, in disagreement with the current LHCb data. On the other hand, the NNPDF40 PDF
generates a large production asymmetry, but it has opposite sign with respect to the data. The inclusion of the recombination process improves the description of the data using the CT18FC parameterization, but still fails to describe the large $p_T$ data.  These results indicate that still exists a room for the improvement of the treatment of the IC component,  as well as of final state effects, in $D^0$ and ${\bar D}^0$ meson production in fixed - target collisions. Surely, more data will be useful to define what  improvements should be performed in future analyses.

\begin{acknowledgments}
V. P. G. would like to thank the members of the Institute of Nuclear Physics Polish Academy of Sciences for their warm hospitality during the completion of this project.
V.P.G. was partially supported by CNPq, FAPERGS and INCT-FNA (Process No. 464898/2014-5). This study was partially supported by the Center for Innovation and Transfer of Natural Sciences and Engineering Knowledge in Rzesz\'ow (Poland).

\end{acknowledgments}

\section*{APPENDIX}
\label{sec:appen}

\subsection{The role of the fragmentation procedure}

In our previous studies \cite{Maciula:2021orz,Maciula:2022otw}, we followed the standard prescription and applied the fragmentation step in the overall center-of-mass system (c.m.s). However, it is known that depending on details it may lead to unstable behavior of the cross-section at rapidities of mesons close to zero and at small transverse momenta. Therefore, here we wish to adopt another procedure, where this problem is removed by performing the fragmentation procedure after the boost to the parton-parton c.m.s. \cite{Szczurek:2020vjn}. As will be shown below, this change has also consequences for the forward/backward regions.

\begin{figure}[t]
\begin{minipage}{0.45\textwidth}
  \centerline{\includegraphics[width=1.0\textwidth]{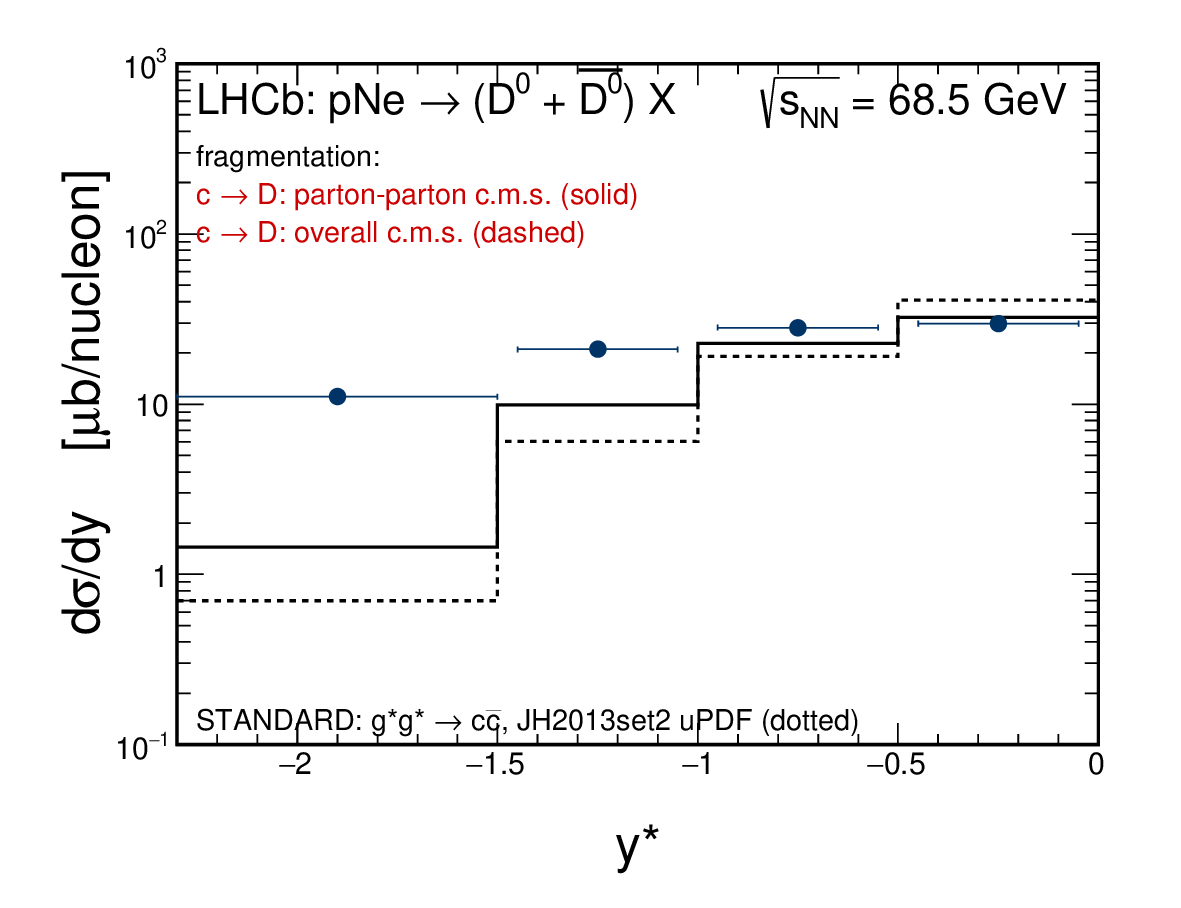}}
\end{minipage}
\begin{minipage}{0.45\textwidth}
  \centerline{\includegraphics[width=1.0\textwidth]{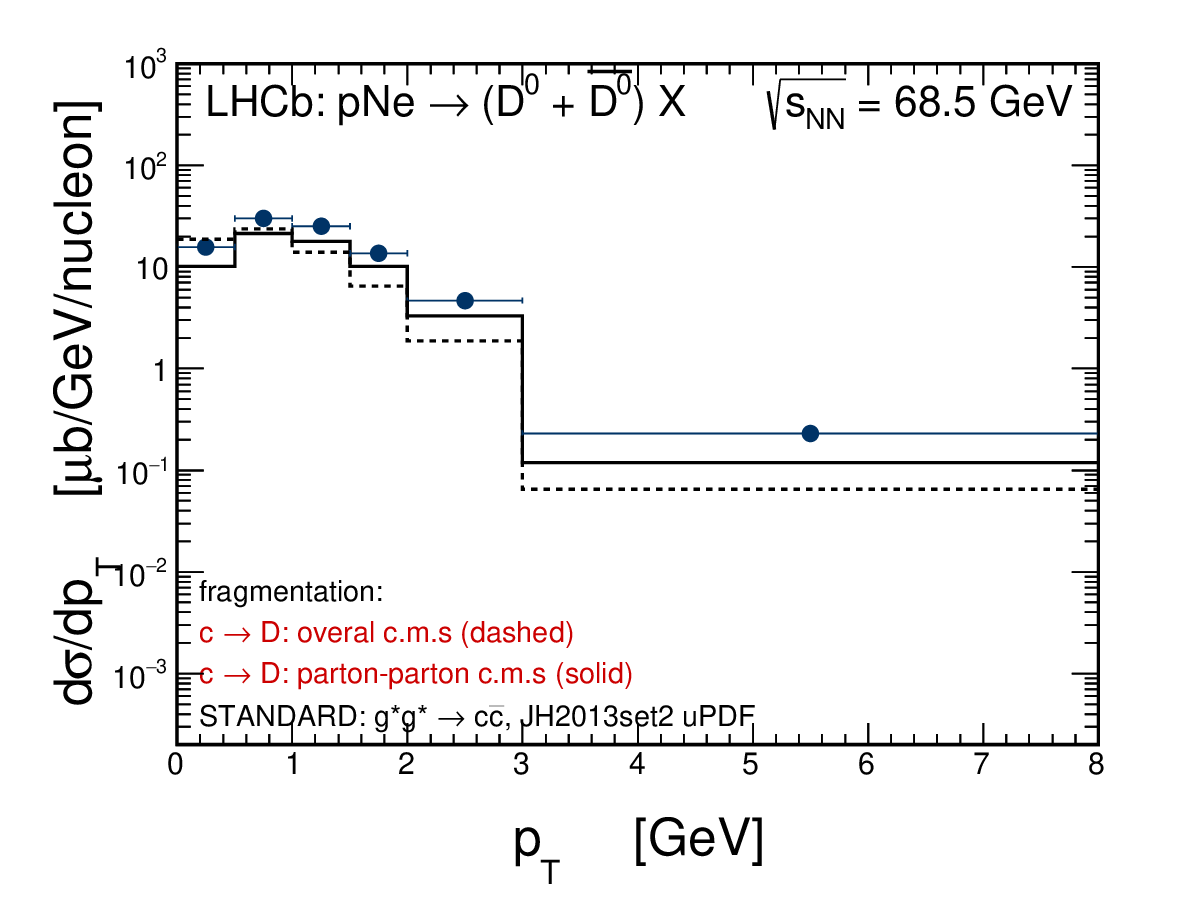}}
\end{minipage}
\begin{minipage}{0.45\textwidth}
  \centerline{\includegraphics[width=1.0\textwidth]{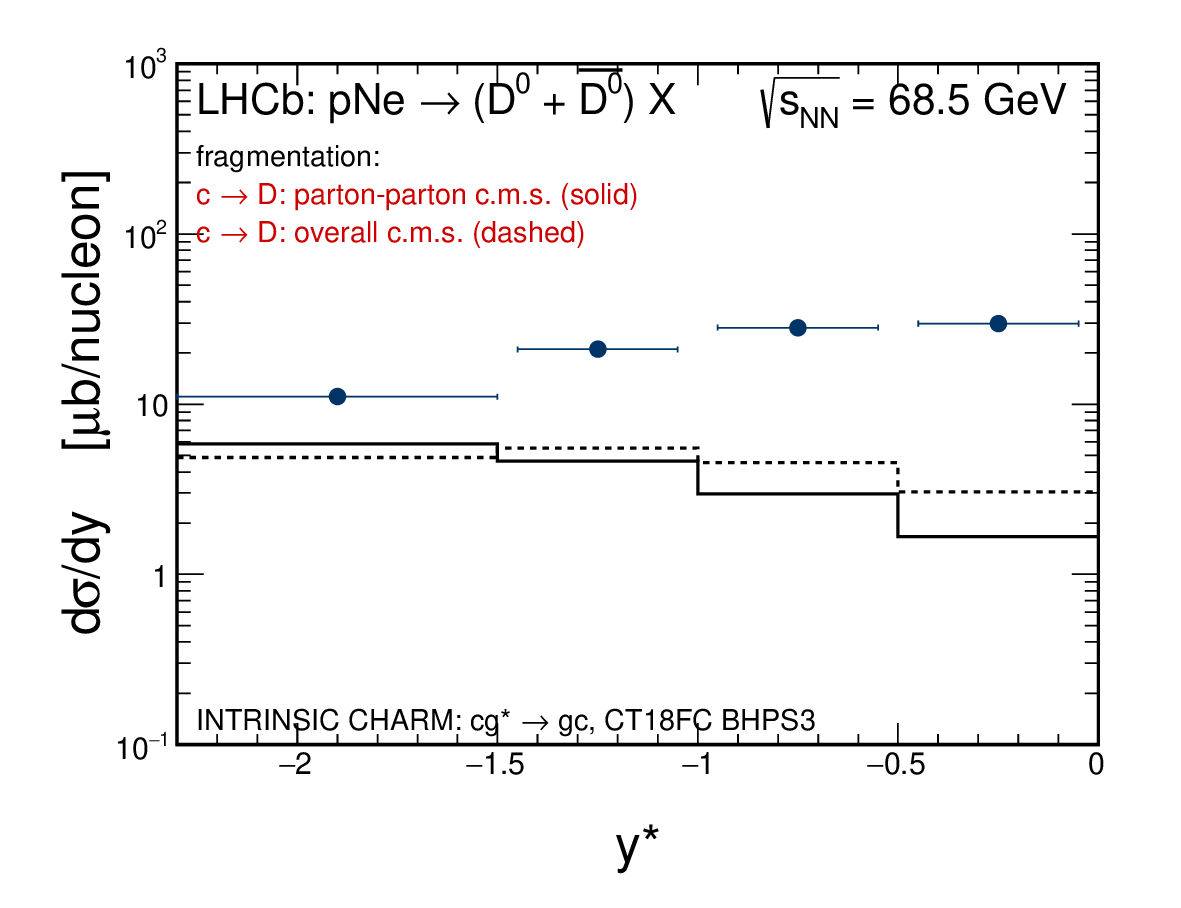}}
\end{minipage}
\begin{minipage}{0.45\textwidth}
  \centerline{\includegraphics[width=1.0\textwidth]{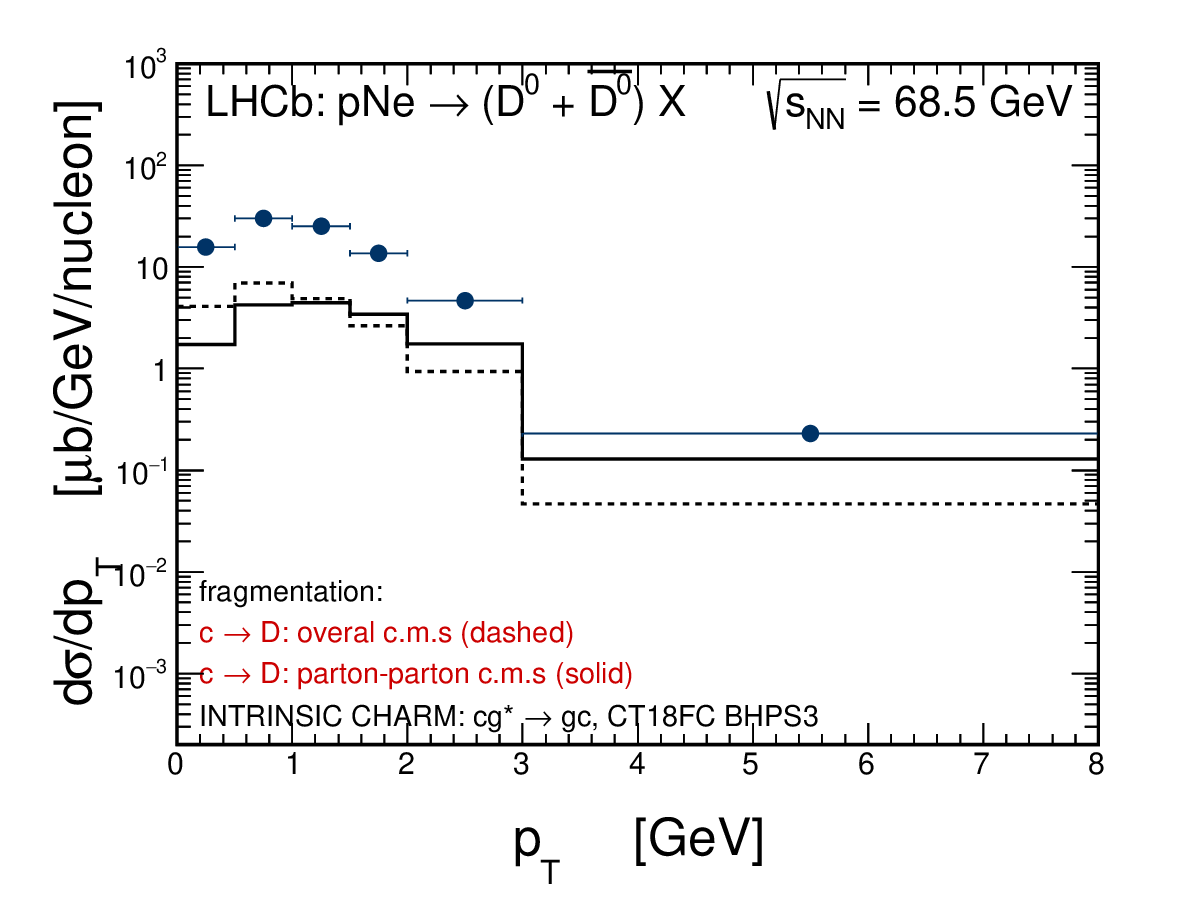}}
\end{minipage}
  \caption{
\small The rapidity (left) and transverse momentum (right) distributions of $D^{0}$ meson (plus $\overline{D^{0}}$ antimeson)
for $p+^{20}\!\mathrm{Ne}$ collisions at $\sqrt{s} = 68.5$ GeV together with the LHCb data \cite{LHCb:2022cul}. Here, we show sensitivity of our predictions for the standard  (upper panels) and gluon - charm (lower panels) mechanisms  to the choice of the fragmentation procedure. Details are specified in the figure. 
}
\label{fig:2}
\end{figure}

In Fig.~\ref{fig:2} we present our predictions for the rapidity (left panel) and the transverse momentum (right panel) distributions of $D^{0}$ meson (plus $\overline{D^{0}}$ antimeson)
for $p\!+\!^{20}\mathrm{Ne}$ collisions at $\sqrt{s} = 68.5$ GeV together with the LHCb data \cite{LHCb:2022cul}. In the upper panels, we present the results considering the standard production mechanism for charm within the $g^*g^* \to c\bar c$ subprocess. The dotted histograms correspond to the fragmentation procedure in the overall c.m.s., while the solid ones to the fragmentation in the parton-parton system. The new procedure leads to larger cross-sections at large meson transverse momenta and in the forward/backward region. On the other hand, at low $p_{T}$'s and for midrapidities it produces slightly smaller results.

A similar comparison, but for the gluon - charm mechanism, $cg^* \to cg$, is shown in the lower panels of Fig.~\ref{fig:2}. The only difference obtained here is that we do not observe an enhancement of the cross-section in the backward rapidity region. Here, the shift of the cross-section in the new fragmentation procedure, with respect to the standard one, appears outside the considered detector acceptance.

%

\subsection{Results for a symmetric intrinsic charm}
\begin{figure}[t]
\begin{minipage}{0.45\textwidth}
  \centerline{\includegraphics[width=1.0\textwidth]{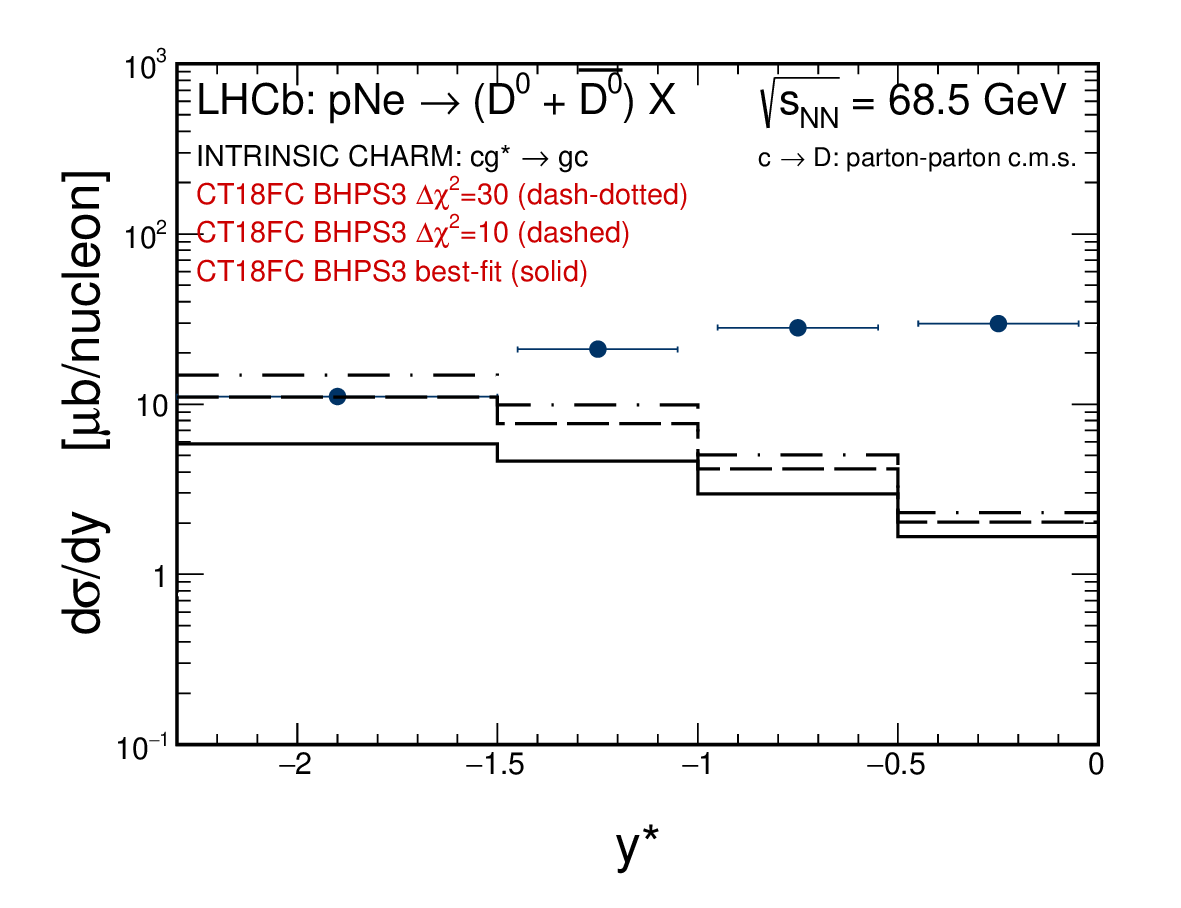}}
\end{minipage}
\begin{minipage}{0.45\textwidth}
  \centerline{\includegraphics[width=1.0\textwidth]{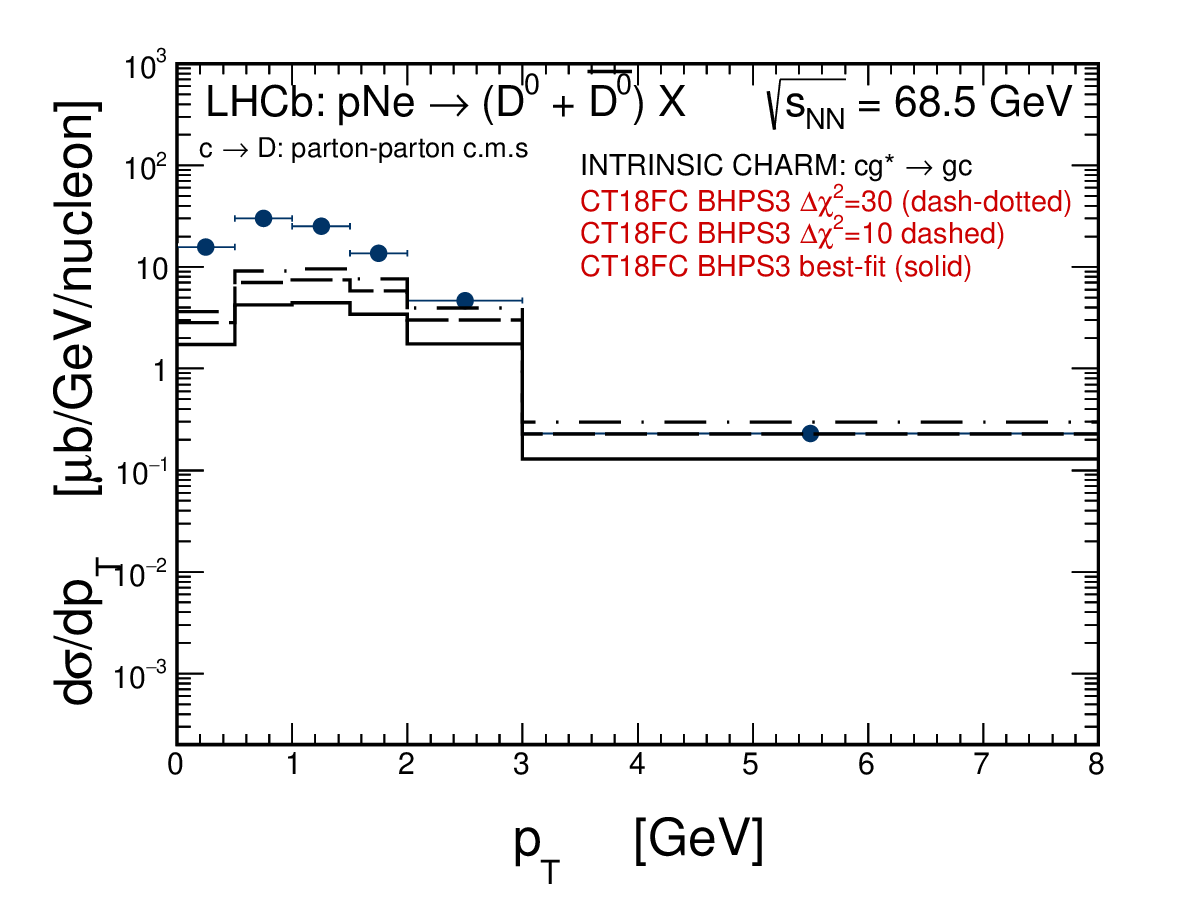}}
\end{minipage}
  \caption{
\small The rapidity (left) and transverse momentum (right) distributions of $D^{0}$ meson (plus $\overline{D^{0}}$ antimeson)
for $p+^{20}\!\mathrm{Ne}$ collisions at $\sqrt{s} = 68.5$ GeV together with the LHCb data \cite{LHCb:2022cul}. Here we show only the BHPS3 intrinsic charm contributions, corresponding to the central (best) fit (solid) and fits at intervals of $\Delta\chi^{2} = 10$ (dashed) and $30$ (dash-dotted).
}
\label{fig:5}
\end{figure}

In addition to the asymmetric charm distributions discussed in the text, the CT18FC parametrization \cite{Guzzi:2022rca} provides also sets which assume $c(x) = \bar{c}(x)$ in the initial condition of the evolution and are based on the BHPS model \cite{Brodsky:1980pb}. In particular, they provide three different grids, which represent different quality of the fits in terms of $\Delta\chi^2$. As a default set in our calculations, we use the set denoted as best-fit. Two other sets correspond to $\Delta\chi^2 = 10$ and $30$. Those different sets estimate uncertainties of the fitted charm PDFs in accord with the common tolerance criteria. In Fig.~\ref{fig:5} we show our results for the mechanism driven by the gluon - charm reaction, calculated within the three different sets of the BHPS3 model. Here, the solid histograms correspond to the central (best) fit, the dashed histograms to the $\Delta\chi^{2} = 10$, and the dash-dotted ones to the $\Delta\chi^{2} = 30$.  The best-fit corresponds to the upper value of the $P_{ic} \lesssim 0.5\%$, while the fit with $\Delta\chi^{2} = 30$ restriction corresponds to the $P_{ic} \lesssim 1.3\%$. These numbers represent a moderate reduction in the allowed upper value of the intrinsic charm probability
with respect to the CT14 IC study\footnote{In the CT14 IC study, the central fit corresponds to  $P_{ic} \lesssim 0.8-1.0\%$, while the upper limit fit corresponds to the $P_{ic} \lesssim 2\%$ \cite{Hou:2017khm}.}.  As expected, a larger value of $P_{ic}$ implies an enhancement of the gluon - charm mechanism, but is not enough to describe the LHCb data.

\begin{figure}[t]
\begin{minipage}{0.45\textwidth}
  \centerline{\includegraphics[width=1.0\textwidth]{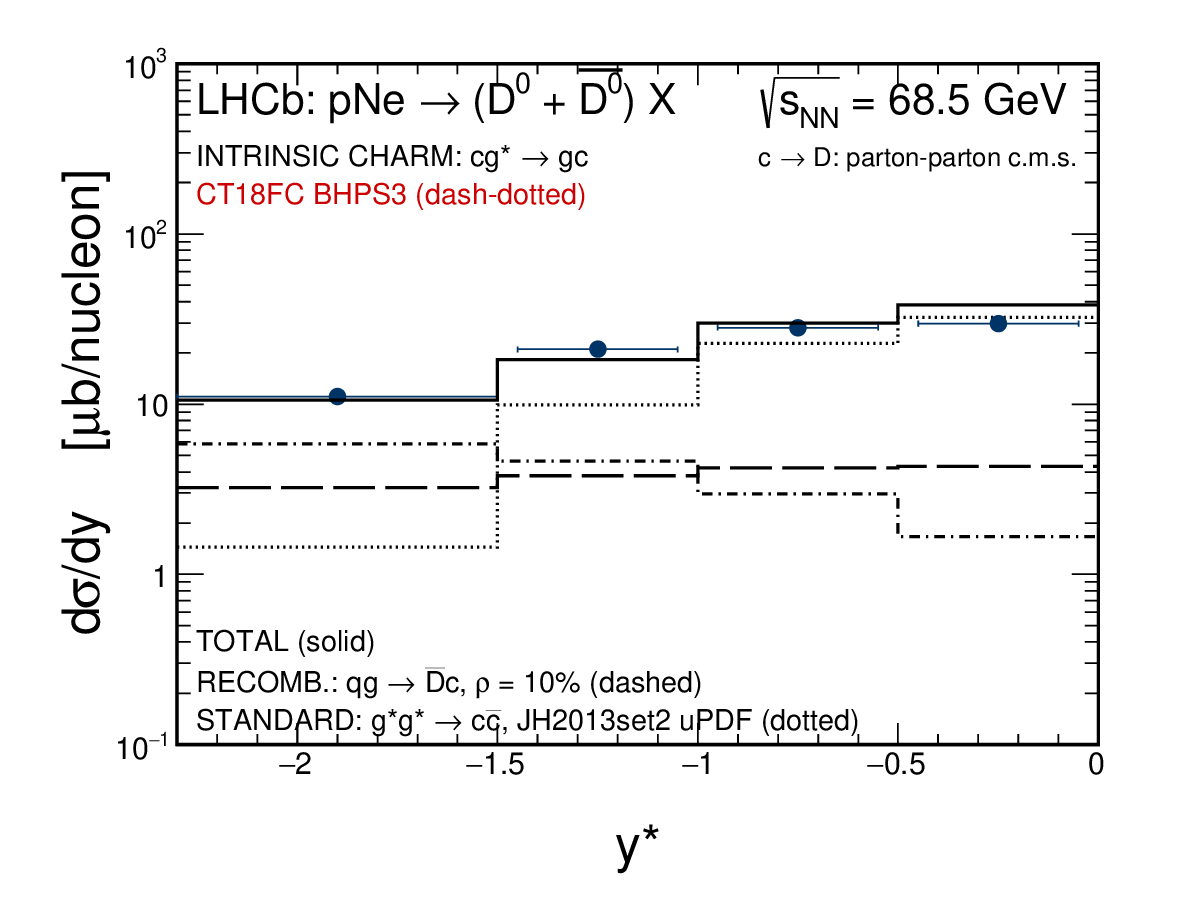}}
\end{minipage}
\begin{minipage}{0.45\textwidth}
  \centerline{\includegraphics[width=1.0\textwidth]{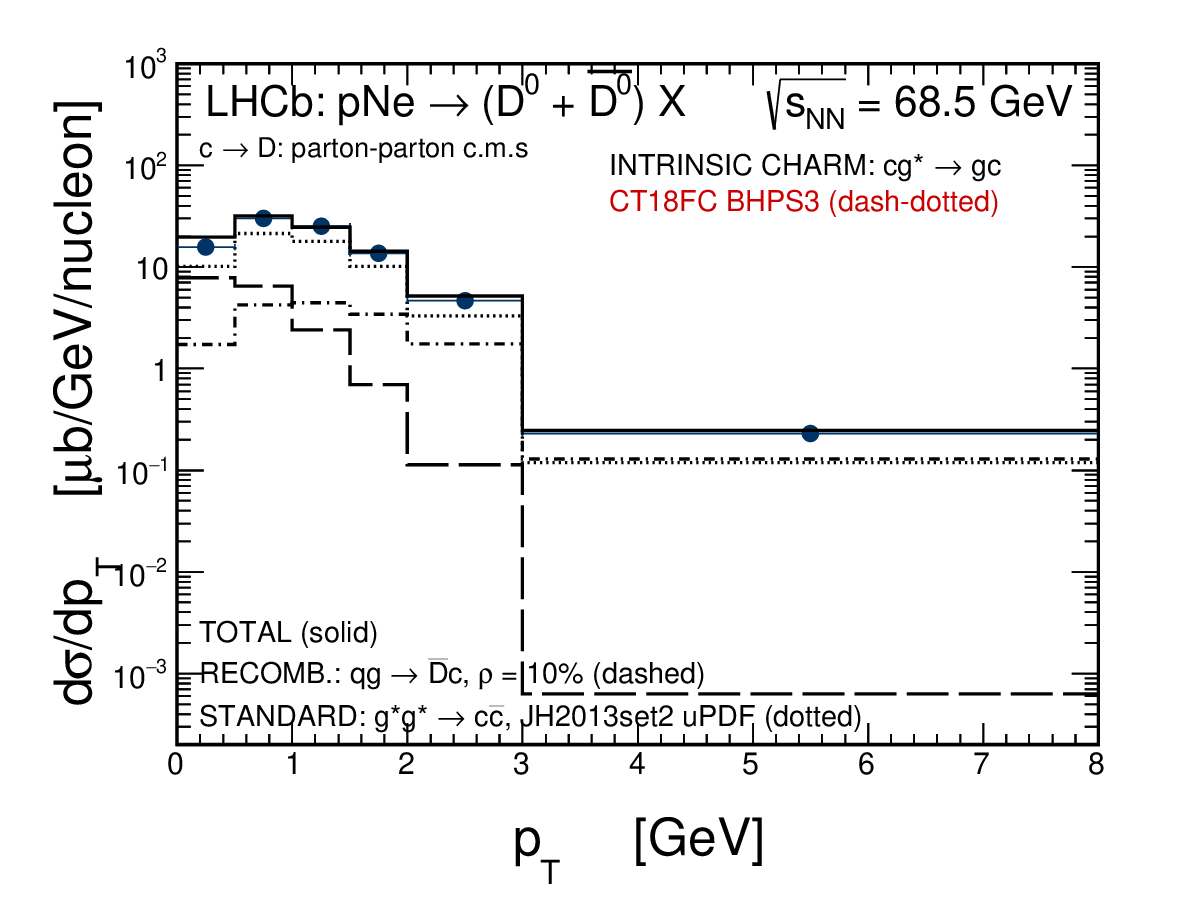}}
\end{minipage}
  \caption{
\small The rapidity (left) and transverse momentum (right) distributions of $D^{0}$ meson (plus $\overline{D^{0}}$ antimeson)
for $p+^{20}\!\mathrm{Ne}$ collisions at $\sqrt{s} = 68.5$ GeV together with the LHCb data \cite{LHCb:2022cul}. Here three different contributions to charm meson production are shown separately, including the standard $g^*g^*\to c\bar c$ mechanism (dotted), the gluon - charm contribution (dot-dashed) and the recombination component (dashed). The solid histograms correspond to the sum of all considered mechanisms. Here, for the intrinsic charm distribution, we use the BHPS3 set of the CT18FC PDF. Details are specified in the figure.
}
\label{fig:4}
\end{figure}

In Fig. \ref{fig:4} we present the predictions for the rapidity and transverse momentum distributions, derived using the central (best) fit of the  BHPS3 set provided by the CT18FC parametrization and taking into account of the gluon - gluon, gluon - charm and recombination processes. As already verified for the asymmetric intrinsic charm distributions, the inclusion of the IC component improves the description of the data. However, it is important to emphasize that if the BHPS3 set is assumed, an $D^{0}$-${\bar D}^{0}$ asymmetry can only be generated by final state effects.



%

\end{document}